\documentclass{elsarticle}
\usepackage[margin=2.5cm]{geometry}
\usepackage[load=named]{siunitx}
\usepackage{color}
\usepackage{verbatim}
\usepackage {caption} 
\usepackage {subcaption}
\usepackage {tabularx} 
\usepackage{arydshln}
\usepackage {hhline}
\usepackage{comment}
\usepackage{amsmath}

\usepackage{lineno}

\usepackage{graphicx}

\makeatletter
\def\ps@pprintTitle{%
 \let\@oddhead\@empty
 \let\@evenhead\@empty
 \def\@oddfoot{}%
 \let\@evenfoot\@oddfoot}
\makeatother

\begin{document}
\begin{frontmatter}

 \title{A multi-physics methodology for four-states of matter}

\author[cav]{L.Michael}
\ead{lm355@cam.ac.uk}

\author[cav]{S.T.~Millmore}
\ead{stm31@cam.ac.uk}

\author[cav]{N.~Nikiforakis}
\ead{nn10005@cam.ac.uk}

\address[cav]{Laboratory for Scientific Computing, Cavendish Laboratory, Department of Physics, University of Cambridge, UK}

\begin{abstract}

  We propose a numerical methodology for the simultaneous numerical
  simulation of four states of matter; gas, liquid, elastoplastic
  solids and plasma. The distinct, interacting physical processes are
  described by a combination of compressible, inert and reactive forms
  of the Euler equations, multiphase equations, elastoplastic
  equations and resistive MHD equations. Combinations of systems of
  equations are usually solved by coupling finite element for solid modelling and CFD
  models for fluid modelling or including material effects through boundary conditions
  rather than full material discretisation. Our simultaneous solution
  methodology lies on the recasting of all the equations in the same,
  hyperbolic form allowing their solution on the same grid with the
  same finite-volume numerical schemes. We use a combination of sharp and diffuse interface methods to track or capture material interfaces, depending on the application.   The communication between the
  distinct systems of equations (i.e., materials separated by sharp interfaces) is facilitated by means of
  mixed-material Riemann solvers at the boundaries of the systems,
  which represent physical material boundaries. To this end we derive
  approximate mixed Riemann solvers for each pair of the above models
  based on characteristic equations. To demonstrate the applicability
  of the new methodology we consider a case study where we investigate
  the possibility of ignition of a combustible gas that lies over a
  liquid in a metal container that is struck by a plasma-arc akin to a
  lightning strike. We study the effect on the ignition of the metal
  container material and conductivity, of the presence of a dielectric
  coating, of insensitive combustible gases and sealed and pre-damaged
  metal surfaces.
\end{abstract}

\end{frontmatter}


\section{Introduction}

The accurate numerical simulation of a wide range of manufacturing, automotive, aerospace and defence processes necessitates the consideration of combinations of gaseous, liquid, elastoplastic solids and plasma. {\color{black}Examples include prevention of ignition sources from entering an aircraft fuel tank containment system and an} explosive in exposed ordnance or mining-sites. Other examples  include additive manufacturing processes such as laser-melt of metal powder.  This article presents numerical methods for the simultaneous solution of equations describing the four distinct states of matter, allowing the non-linear communication of the materials in a multi-physics framework.  The methodology can also be used as part of the manufacturing process for optimising industrial (e.g., plane-wing coatings, car  etc.) components in terms of shapes and materials.

An integrated numerical methodology for numerically simulating an application involving four states of matter has three elements; the formulations describing each state of matter, the numerical methods that solve these separately and the communication between any two models through material coupling. Here we give an overview of relevant literature concerning the first and the last of these elements, as this is where the novelty of this work lies. In the rest of this section we summarise the background literature of formulations modelling elastoplastic solids, plasma and fluids (liquids and gases) and give an overview of existing communication methods between these.

The mathematical description of the elastoplastic system has been traditionally done in a Lagrangian framework. The original Lagrangian form of the solid equations has been reformulated into a conservative form of equations in the Eulerian frame by Godunov and Romenskii \cite{godunov1972nonstationary}, Kondaurov \cite{kondaurov1982equations} and Plohr and Sharp \cite{plohr1988conservative}. This has the advantage of allowing the solution of the elastoplastic solid formulation in the same framework as the  hydrodynamic formulation for fluids, using the same (or the same family of) high-resolution, shock-capturing methods. This led to the development of high-order, shock capturing schemes for the numerical solution of such systems. For example, Miller and Colella \cite{miller2001high} and Barton et al.\ \cite{barton2009exact,barton2010plastic} have developed linearised Riemann solvers as part of a high-order numerical scheme to capture the seven waves in the (1D) solid system, while Gavrilyuk et al.\ \cite{gavrilyuk2008modelling} have presented the adaptation of the classic HLLC solver to the solid system. Centred numerical schemes and linearised Riemann solvers for the solid systems have also been presented by Titarev \cite{titarev2008musta}, while approximate and exact Riemann solvers for the conservative elastic system have been presented by Miller \cite{miller2004iterative} and Barton et al.\ \cite{barton2009exact}.

Inclusion of plasticity in the solid system has been presented using different approaches, as for example by Miller and Colella \cite{miller2001high}, who evolve the plastic deformation gradient ($\mathbf{F}^p$) in addition the total inverse deformation gradient ($\mathbf{G}=\mathbf{F}^{-1}$) and include an elastic predictor step followed by a `plastic' corrector step to correct any over-estimated elastic deformation that pushes the state outside the yield surface. The predictor-corrector approach allows for solving both for perfect and time-dependent plasticity models. Another approach is followed by Barton et al.\ \cite{barton2010plastic} who only evolve the elastic deformation gradient ($\mathbf{F}^e$) and include plasticity as source terms for the elastic deformation tensor equations.

The full system for modelling resistive ionised material in the presence of electric and magnetic fields is the resistive MHD equations, which couple the Navier-Stokes equations to Maxwell'’s equations. If these are used to model lightning strikes, or laboratory based plasma arcs, they couple vastly different timescales; the electromagnetic effects have a characteristic timescale of the speed of light, whilst the mechanical dynamics of the system have a characteristic timescale of the speed of sound. Interesting numerical studies on lightning modelling include work by Plooster \cite{plooster1971numerical,plooster1971spark,plooster1970shock}, Paxton et al.\ \cite{paxton1986lightning}, Aleksandrov et al.\ \cite{aleksandrov2000effect}, Tanaka et al.\ \cite{tanaka2005hydrodynamic}, Chemartin \cite{chemartin2009three,chemartin2011modelling,chemartin2012direct}, Villa \cite{villa2011multiscale} and Tholin \cite{tholin2015numerical}. 

The mathematical description of the fluids (liquid and gas) could vary in complexity, depending on the physical properties of the material that are dominant in the application of consideration.  Formulations relevant to this work{\footnote{that can be used in our multi-physics methodology} include members of the Euler and augmented Euler class of conservation laws (e.g., \cite{banks2007high,banks2008study,wang2004thermodynamically,shyue1998efficient,shyue1999fluid,shyue2001fluid}), the BN-type and reduced model class of conservation laws  (e.g., \cite{baer1986two,saurel1999multiphase,kapila2001two,murrone2005five}) or hybrid formulations (e.g., \cite{michael2016hybrid,schwendeman2012hybrid}).

Traditionally, solid-fluid coupling is done following Lagrangian techniques. These, however, present difficulties for large deformations of the materials as these are inherently translated to large deformations in the underlying mesh. Others methods for solid-fluid coupling include Smoothed Particle Hydrodynamics (SPH) and arbitrary Lagrangian-Eulerian (ALE). Also, in solid-fluid problems it is common to use one-way coupling of finite element codes describing the solid material with CFD codes modelling the fluid part is used. As a result, the two processes are solved in a `co-simulation' environment, each on it's own grid with a distinct numerical method, without the non-linear interaction of the materials involved. This may also lead to discretisation errors passing from one method to the other. Studies
including solid-fluid coupling in the Eulerian frame include work by Miller
and Colella \cite{miller2002conservative} and Barton et al.\ \cite{barton2011conservative},
Schoch et al.\ \cite{schoch2013eulerian}, Monasse \cite{monasse} in an embedded
boundary framework, Favrie et al.\ \cite{favrie2009solid,favrie2012diffuse}
in a diffuse interface approach and a combination of sharp and diffuse interface
approach by Michael and Nikiforakis \cite{michael2018multi}.
To the authors' knowledge, the simultaneous solution of plasma-arc and a solid has not been presented before.  The interaction of the lightning arc with a surface (e.g, \ \cite{chemartin2009three,chemartin2011modelling,chemartin2012direct,chemartin2008modelisation,lago2004modelisation,ogasawara2010coupled,abdelal2014nonlinear,guo2017comparison,foster2018understanding,karch2015contributions}) is usually done by considering  prescribed plasma or prescribed material conductivity at the material interface rather than considering full solution of the two phases simultaneously.  

The literature suggests that previous work or existing commercial codes used for multi-material and multi-phase applications can often only handle limited (less than four) combinations of states of matter. For lack of better methods, the trend is to do so using a combination of different codes in a `co-simulation' environment (e.g.,  the fuel interaction with the tank walls) or include material effects through boundary conditions (e.g., lightning striking a metal substrate). For some material combinations (e.g, lightning striking a metal), it has not even been possible so far to solve for the two phases (plasma arc, elastoplastic solid) simultaneously.

In this work, the methodology for the simultaneous modelling of four states of matter in a single computational code and a single configuration is presented for the first time. It consists of three components: i) casting the governing equations in a hyperbolic form, ii) which in turn facilitates the solution of each system using the same, high-resolution, shock-capturing numerical methods, on the same grid and iii) appropriately solving mixed-material Riemann problems at material interfaces to achieve the communication between  materials. Moreover, the innovative combination of diffuse and sharp interface methods allowing the non-linear coupling of the four states of matter is also presented here. We also demonstrate through a case study the capability of the methodology to solve simultaneously for a plasma arc and  elastoplastic solids, in an application involving four states of matter; the lighting strike on a metal container, half-filled with liquid and half-filled with combustible gas. 

 To describe the elastic behaviour of the solid material in this work, we use the elastic deformation evolution model by Barton et
al.\ \cite{barton2009exact}, further extended later in the work by Schoch et al. \cite{schoch2013eulerian,schoch2013propagation} and Michael and Nikiforakis \cite{michael2018multi}. Inelastic deformation is following the approach of Miller and Colella of a predictor-corrector method based on the principle of maximum dissipation and  is applied in combination with perfect plasticity and time-dependent plasticity models. Regarding the plasma modelling, resolving the
entire resistive MHD system would be inefficient and unnecessary for the lightning context. Thus, in this work we reduce
the MHD equations to mechanical equations which interact with the electrodynamic
fields, following \cite{michael2018numerical,millmore2019,villa2011multiscale,tholin2015numerical}. The fluids in this work are described by the compressible form of the Euler equations or multi-phase models suitable for gas, liquid and explosive modelling. Our methodology does not restrict us to a particular formulation for fluid modelling and all formulations in the spectrum of Euler to BN-type and hybrid formulations can be used. However, here we choose to use, based on the application for the case study, the compressible Euler equations for the liquid modelling and augment the plasma model with an equation that allows the multi-phase transition between gas and plasma via the equation of state.

The complete multi-physics system is represented
in an Eulerian frame, on a regular Cartesian mesh using Cartesian data structures and all systems of equations are solved with finite
volume techniques, employing high-resolution, shock-capturing methods (e.g.,
MUSCL Hancock with HLLC). The communication between the different systems
is achieved by employing the Riemann ghost fluid method and the mixed-material
Riemann solvers presented here.  

In the remainder of this article, we first present the distinct mathematical formulations describing each of the four states of matter. The technique for communication between the different phases is then presented including the  mixed Riemann solvers for plasma-fluid and plasma-solid pairs. The communication between the remaining pairs of  states of matter can either be found in \cite{michael2018multi} or can be trivially extracted from these. Finally, we present the case study where we investigate the possibility of ignition of a combustible gas, lying over a liquid in a metal container that is struck by a plasma-arc. We investigate the effect on {\color{black}ignition prevention} of the material of the metal container (using aluminium and low-conductivity materials), the effect of the presence of a dielectric coating, sealed and pre-damaged metal containers and insensitive combustible gases.

\section{Mathematical models}
\label{models}
In this section, the distinct mathematical formulations describing the
four states of matter are presented. The key behind this step is
writing all systems of equations in the same, hyperbolic mathematical
form, given in one dimension as :

\begin{equation}
\mathbf{U}_t+\nabla \cdot {\bf F}=\mathbf{S(U)}.
\end{equation}

\subsection{Phases 1 and 2: Modelling the Plasma and the Gas}
\label{plasma}
 In this work, we describe the evolution of a plasma
  arc using the MHD approach of Chemartin et
    al.\ \cite{chemartin2009three}.  Over the timescales we consider,
  viscosity effects are negligible and we reduce the system of
  evolution equations to three conservation laws for the mechanical
  properties of the plasma; density, $\rho$, momentum, $\rho v$, and total energy,
$\rho E$:

\begin{eqnarray}
\label{Villa:cty}
\frac{\partial \rho}{\partial t}+ \nabla \cdot(\rho \mathbf{u})  &=&0,\\
\label{Villa:mom}
\frac{\partial}{\partial t}(\rho \mathbf{u_i})+\nabla \cdot(\rho \mathbf{u_iu})+\frac{\partial
p}{\partial{\mathbf{x_i}}}&=&\mathbf{J}\times\mathbf{B},\\
\label{Villa:ene}
\frac{\partial}{\partial t}(\rho E)+\nabla \cdot (\rho E+p)\mathbf{u}&=&  \mathbf{u} \cdot (\mathbf{J} \times \mathbf{B})+ \eta \mathbf{J}\cdot\mathbf{J} + S_r+KQ_{\text{det}}, \\
\label{Villalambda}
\frac{\partial \rho \lambda}{\partial t}+ \nabla \cdot(\rho \mathbf{u}\lambda)  &=&\rho K,\\
 \label{Villa:magn}
-\nabla^2\mathbf{A}&=& \mu_0\mathbf{J},\\
\label{Villa:current}
\nabla \cdot \mathbf{J} &=& -\nabla \cdot(\sigma \nabla \phi)=0,
\end{eqnarray}
where $p$ is the pressure, $\mathbf{J}$ the current density,
$\mathbf{B}$ the magnetic field, which is related to the magnetic
vector potential $\mathbf{A}$ by
$\mathbf{B} = \nabla \times \mathbf{A}$, $\eta$ the resistivity of the
plasma, $S_r$ a term for the radiative losses from a heated material,
$\sigma=1/\eta$ is the electrical conductivity and $\phi$ is the
electric potential. The current density is governed by the boundaries
of the domain of interest; for simulations of laboratory arc
attachment experiments, there will be an input current and a ground
site associated with the circuitry of the experiment.

To model the gaseous explosive, we use the above system of equations,
with the hydrodynamic evolution described by
(\ref{Villa:cty})-(\ref{Villa:ene}) and the combustion process by the
equation (\ref{Villalambda}) and the source term $KQ_{\text{det}}$ in
(\ref{Villa:ene}).  In (\ref{Villalambda}), we evolve $\lambda$, the
reaction progress variable, taking values between 0 and 1, where 1
indicates that no material has reacted yet and 0 that all material has
reacted. The term $KQ_{\text{det}}$ in (\ref{Villa:ene}) represents
the energy added to the system due to the combustion. The combustion
process is assumed to follow a single-step Arrhenius law, such that
\begin{equation}
\label{arrhenius}
\frac{d\lambda}{dt}=K=-\lambda Ce^{-\frac{T}{T_A}},
\end{equation}
where $T_A$ is the gas activation temperature, $T$ the local temperature of the material and $C$ a pre-exponential factor given for the specific material, describing how fast the reaction grows. The heat of detonation is denoted by $Q_{\text{det}}$. 

\subsubsection{Current density solution}
\label{CDS}

In order to allow for the full interaction of a plasma arc with a
substrate of different conductivities, the full current flow through
the substrate has to be computed, as described later in the
elastoplastic solids modelling section.  By including the interaction
of the arc plasma with the substrate, we expect to lose the symmetry
in the z- direction of a steady arc. As a result, we expect a radial
component to the current density profile (restricting this work to
cylindrical symmetry, the azimuthal component remains
zero). Effectively, this gives three elliptic equations to solve, the
current continuity equation,
\begin{equation}
\label{sigmaequ}
\nabla \cdot(\sigma\nabla\phi)=0
\end{equation}
and the $r$- and $\theta$-components of the magnetic potential vector
equation (\ref{Villa:magn}).

  When solving these elliptic equations, challenges can arise when
  there are discontinuous changes in conductivity within the domain;
  particularly where the edge of the substrate is adjacent to
  unionised air, conductivity can drop from
  $\mathcal{O} (10^7) \SI{}{\siemens \per \meter}$ to zero. These
  challenges arise due to the fact that derivative quantities (e.g.,
  $\nabla \sigma$ ) are undefined at these discontinuities. Instead of
  solving the equations in this form, one common technique is to cast
  them into an integral (weak) form, and apply solution techniques to
  this form.

\subsubsection{Equation of state}
The system of equations (\ref{Villa:cty})-(\ref{Villa:magn}) comprises
9 equations for 10 unknown variables, $\rho,E,p, \lambda$ and the
vectors $\mathbf{u}$ and $\mathbf{B}$. These are closed through an
equation of state, which describes the thermodynamic properties of the
system, typically written in the form $p = p (e, \rho)$, where $e$ is
the specific internal energy, and is related to the specific total
energy through $\rho E = \rho e + \frac{1}{2}u^2$ . The equation of
state of a plasma is complex, since its thermodynamic properties
depend strongly on the degree of ionisation.

 Over the range of temperatures we consider in this
  work, from ambient air to a plasma arc centre of over 50,000\,K, the
  ratio of specific heats changes considerably, and hence we cannot
  use a simple ideal gas-based model.  Computing accurate thermodynamic
  properties then requires detailed knowledge of the composition of a
  plasma for a given thermodynamic state (e.g., at a given density and
  internal energy).  Here we use a tabulated equation of state based on
  the 11-species model of Mottura et al.\ \cite{mottura1997evaluation},
  describing the plasma properties over a range of pressure and
  temperatures. This computes the contribution to energy, sound speed
  and density from the various species which comprise the
  plasma. Electrical conductivity is provided as a function of
  composition following D'Angola et al.\ \cite{d2008thermodynamic}. The
  behaviour of the equation of state as an ideal gas at low temperatures
  allows the simultaneous modelling of gas and plasma, assuming the gas
  follows an ideal gas law.

\subsubsection{Pre-heated arc region}

Our current density solution technique incorporates an electrode for
current input within the domain; the initial data for this work
require a `pre-heated' arc region at the centre of the
domain. Physically, the onset of the arc connection is a voltage
breakdown of the air between the electrode and the substrate, which
generates a conductive path for the current to take. This is a complex
process which  occurs over timescales much shorter
  than the plasma arc evolution, and is still not fully understood,
and hence cannot be incorporated within our
numerical model. The standard technique to provide a conductive path
for the current is to assume a thin, high-temperature domain at the
centre of the arc
\cite{tholin2015numerical,chemartin2009three,larsson2000lightning}. It
has been shown by Larsson et al.\ \cite{larsson2000lightning} that
this does not affect the overall behaviour of the arc. We use a
preheated region of $\SI{8000}{\kelvin}$ with a
$2\SI[load=text]{}{\milli \metre}$ diameter. The current profile used
in this work is taken from the literature \cite{arp5412b} (Component A
theirin) and has a rapid ride to a maximum current of
$\SI{200}{\kilo \ampere}$ over $\SI{6.4}{\micro \second}$ and is
determined by:

\begin{equation}
I(t) = I_0(e^{-\alpha t}-e^{-\beta t})(1-e^{-\gamma t})^2,
\end{equation}
where $I_0=218,810\SI{}{\ampere}, \alpha = 11,354\SI{}{\per \second},
\beta = 647,265\SI{}{\per \second} \; \text{and} \;  \gamma =
5,423,540\SI{}{\per \second}$. 

\subsection{Phase 3: Modelling liquids}
\label{gas}
To describe the behaviour of a liquid in this context of work, the compressible
form of the Euler equations suffices\footnote{We do not see sufficient temperature rise to justify ionisation and require consideration of MHD in this fluid part.}:

\begin{eqnarray}
\label{Euler:cty}
\frac{\partial \rho}{\partial t}+ \nabla \cdot(\rho \mathbf{u})  &=&0,\\
\label{Euler:mom}
\frac{\partial}{\partial t}(\rho \mathbf{u_i})+\nabla \cdot(\rho \mathbf{u_iu})+\frac{\partial
p}{\partial{\mathbf{x_i}}}&=&0,\\
\label{Euler:ene}
\frac{\partial}{\partial t}(\rho E)+\nabla \cdot (\rho E+p)\mathbf{u}&=&0,
\end{eqnarray}
where $\mathbf{u}=(u,v,w)$ denotes the total vector velocity, $i$ denotes
space dimension, $i=1,2,3$, $\rho$ the total density of the fluid and $E$
the specific total energy given by $E=e+\frac{1}{2}\sum_{i}u_i^2$, with $e$
the total specific internal energy of the system. This form of equations
can be used to describe also other gases, especially ones that do not follow
the ideal gas law and hence are not modelled by the low-temperature regime
of the plasma equation of state.

The particular fluid (gas or liquid) considered is determined by the equation
of state that closes the system. For example, air can be modelled by the
ideal gas law ($p = \rho(\gamma -1)e$), and water by the stiffened gas equation
of state ($p = \rho\Gamma-\gamma p_{\infty}$).

More complex materials, such as reactive components, can be modelled by the MiNi16
model \cite{michael2016hybrid} or the MiNi16-reduced formulations and more complex equations of state like the JWL or Cochran-Chan. In the
case of explosive modelling, we could use the MiNi16 model or its reduced
versions to model the combustion of the material, in which we could have
one liquid/solid reactive material turning into gaseous reactants.

\subsection{Phase 4: Modelling elastoplastic solids}
\label{solids}
In this work, we use the elastic solid model described by Barton et al.\ \cite{barton2009exact}, Schoch
et al. \cite{schoch2013eulerian,schoch2013propagation} and Michael and Nikiforakis
\cite{michael2018multi},
based on the formulation by Godunov and Romenskii \cite{godunov2013elements}.
The plasticity is included following the work of Miller and Colella \cite{miller2002conservative}.

 If we consider an elastoplastic material in
isolation, then in an Eulerian frame, which we employ here,
  there is no mesh distortion that can be used to describe the solid
  material deformation. The material distortion
   therefore needs to be
  accounted for in a different way. Here, this is done by defining the
  total deformation gradient tensor as:
\begin{equation}
\label{Fe}
F_{ij} = \frac{\partial x_i}{\partial X_j},
\end{equation}
which maps the initial configuration (coordinate $\mathbf{X}$)
to the deformed configuration (coordinate $\mathbf{x}$). Following the proposition
by Rice \cite{rice1971inelastic} we introduce an intermediate configuration
describing only plastic deformations described by $\mathbf{F}^p$ such that
the total deformation is decomposed multiplicatively into elastic and plastic
components as $\mathbf{F}=\mathbf{F}^e\mathbf{F}^p$.

The state of the solid is characterised by the elastic deformation gradient,
velocity $u_i$ and entropy $S$. Following the work by Barton et al.\ \cite{barton2009exact},
the complete three-dimensional system forms a hyperbolic system of conservation
laws (\ref{cty})-(\ref{materialhistory}) for momentum, strain and energy:

\begin{eqnarray} \label{cty}
  \frac{\partial{\rho u_i}}{\partial t} + \frac{\partial(\rho u_i u_m - \sigma_{im})}{\partial
x_m} &=& 0, \\
  \frac{\partial{\rho E}}{\partial t} + \frac{\partial(\rho u_m E - u_i \sigma_{im})}{\partial
x_m} &=& \eta J_i J_i, \\
\frac{\partial{\rho F^e_{ij}}}{\partial t} + \frac{\partial(\rho F^e_{ij}u_m-\rho
F^e_{m j}u_i)}{\partial x_m} &=&  -u_i\frac{\partial \rho F^{e}_{mj}}{\partial
x_m} + P_{ij}, \label{F}\\
  \frac{\partial{\rho \kappa}}{\partial t} + \frac{\partial(\rho u_m
    \kappa)}{\partial x_m} &=& \rho \dot{\kappa},\label{materialhistory}\\
-\nabla^2A_i &=& \mu_0 J_i,
\end{eqnarray}
with the vector components $\cdot_i$ and tensor components
$\cdot_{ij}$. The first two equations along with the density-deformation
gradient relation:
\begin{equation}
\rho = \rho_0 / \text{det}\mathbf{F}^e,
\end{equation}
where $\rho_0$ is the density of the initial unstressed medium,  essentially
evolve the solid material hydrodynamically. Here, $\sigma$ is the stress,
$E$ the specific total energy such that $E=\frac{1}{2}|u|^2+e$, with $e$
the specific internal energy and $\kappa$ the scalar material history that
tracks the work hardening of the material through plastic deformation.
We denote the source terms associated with the plastic update as $P_{ij}$.

Electrodynamic effects are present within the solid material, though only
through the Joule effect in the energy equation ($\eta J_i J_i$). The Lorentz
effect does not apply to solid materials, where intermolecular forces prevent
material flow with the magnetic field. We therefore solve for the magnetic
vector potential  and current density as we do for a
plasma  through equations (\ref{Villa:magn}) and (\ref{Villa:current}).

The system is closed by an analytic constitutive model relating the specific
internal energy to the deformation gradient, entropy and material history
parameter (if applicable):
\begin{equation}
e = e(\mathbf{F}^e,S,\kappa).
\end{equation}
Restricting this work to isotropic materials, $e(\mathbf{F}^e)=e(I_1,I_2,I_3)$, with $I_1,I_2,I_3$ are the invariants of the Finger tensor given by $ \boldsymbol{C} = \boldsymbol{F}^{-T}\boldsymbol{F}$.
The stress tensor is given by:
\begin{equation}
\sigma_{ij} = \rho F^e_{im}\frac{\partial e}{\partial F^e_{jm}} 
\end{equation}
or equivalently, 
\begin{equation}
\boldsymbol{\sigma} = -2\rho \left[\frac{\partial\mathcal{E}}{\partial I_3}I_3\ \mathcal{I} + \left(\frac{\partial\mathcal{E}}{\partial I_1} + I_1 \frac{\partial\mathcal{E}}{\partial I_2}\right) \boldsymbol{C} - \frac{\partial\mathcal{E}}{\partial I_2} \boldsymbol{C}^2\right].
\end{equation}

For $\mathbf{F}$ to represent physical, continuous deformations,
it must satisfy three compatibility constraints:
\begin{equation}
\frac{\partial \rho F_{kj}}{\partial x_k} = 0, \quad j=1,2,3,
\end{equation}
which hold true for $t>0$ if true for initial data. This is based on the
fact that $\mathbf{F}$ is defined as a gradient. 

The deformation is purely elastic until the stress state is evolved beyond
the yield surface ($f>0$), which in this work is given by the Von Mises criterion:

\begin{equation}
\label{yieldSurface}
f(\boldsymbol{\sigma} )=||\text{dev}\boldsymbol{\sigma}||-\sqrt{\frac{2}{3}}\sigma_Y
= 0, \; \text{with} \;\; \text{dev} \boldsymbol{\sigma} =  \boldsymbol{\sigma}-\frac{1}{3}(\text{tr}\boldsymbol{\sigma})I,
\end{equation}
where $\sigma_Y$ is the yield stress and the matrix norm $||.||$ the Shur
norm ($||\boldsymbol{\sigma}||^2=\text{tr}(\boldsymbol{\sigma}^T\boldsymbol{\sigma}$)).

As this identifies the maximum yield allowed to be reached by an elastic-only
step, a predictor-corrector method is followed to re-map the
solid state onto the yield surface. Assuming that the simulation timestep
is small, this is taken to be a straight line, using the associative flow
rate 
($\dot{\epsilon}^p=\eta\frac{\partial F}{\partial \sigma}$), satisfying the
maximum plastic dissipation principle (i.e.\ the steepest path). In general,
this is re-mapping procedure is governed by the dissipation law: 
\begin{equation}
  \psi_{plast} =\boldsymbol{\Sigma} \colon ((\mathbf{F}^p)^{-1}\dot{\mathbf{F}^p}),
\end{equation}
where $\boldsymbol{\Sigma}=\boldsymbol{G}\boldsymbol{\sigma}\boldsymbol{F}$
and $\colon$ is the double contraction of tensors (e.g., $ \boldsymbol{\sigma}:\boldsymbol{\sigma}=\text{tr}(\boldsymbol{\sigma}^T\boldsymbol{\sigma})$).
The initial prediction is $\mathbf{F} = \mathbf{F}^e$ and 
$\mathbf{F}^p = \mathbf{I}$, where $\mathbf{F}$ is the specific total deformation
tensor and $\mathbf{F}^p$ the plastic deformation tensor that contains the
contribution from plastic deformation. This is then relaxed to the yield
surface according to the procedure of Miller and Colella  \cite{miller2002conservative}.

For the aluminium used in this work, we use the Romenskii hyperelastic equation
of state \cite{titarev2008musta}, consisting of a two-term hydrodynamic component
and a one-term shear deformation component. The full equation is given by:
\begin{equation}
\epsilon(I_1,I_2,I_3,S) = \frac{K_0}{2\alpha^2}\left(I_3^{\alpha/2}-1\right)^2+
c_v T_0 I_3^{\gamma/2}\left(e^{S/c_v}-1\right) +\frac{B_0}{2}I_3^{\beta/2}\left(\frac{I_1^2}{3}-I_2\right),
\label{RomenskiiElastic}
\end{equation}
where $I_1,I_2,I_3$ are the invariants of the Finger tensor, $K_0=c_0^2 -
\frac{4}{3}b_0^2$ is the squared bulk speed of sound, $B_0$ is the reference
shear wave speed, $c_v$ is the specific heat capacity at constant volume,
$T_0$ is the reference temperature, $\alpha$ and $\gamma$ are exponents determining
the non-linear dependence of the sound speed and temperature on density respectively
and $\beta$ an exponent determining the non-linear dependence of this shear
wave speed on density.

The constitutive model parameters for the aluminium considered here are given
in Table \ref{eosTable}. Perfect plasticity is assumed, with a yield stress
of $\SI{0.4}{\giga \pascal}$. The conductivity of the aluminium is taken
to be $3.2\times10^7\SI{}{\siemens \per \meter }$.

\begin{table}[!t]
\centering
 \begin{tabular}{c c c c c c c c c}\hline 
\textbf{Hyperelastic and} & $\rho_0$   & $c_v$   & $T_0$  & $\alpha$ & $\Gamma_0$
& $b_0$  & $c_0$  & $\beta$ \\ 
\textbf{shear parameters} & [\SI{}{\kilogram \per \meter \tothe{3}}] &  [\SI{}{\joule
\per \kilogram \per \kelvin}] &  [\SI{}{\kelvin}] & - & - &  $[\SI{}{\meter
\per \second}]$ & $[\SI{}{\meter \per \second}]$ & - \\ \hline 

Aluminium / CFRP & 2710 & 900 & 300 &  1 &   2.088 &  3160 &  6220 &  3.577
 \\\hline
\end{tabular}
\caption {Romenskii equation of state parameters the elastoplastic solid
materials used in this work.}
\label{eosTable}
\end{table}

For the PMMA dielectric considered in this work we use a shock Mie-Gr\"uneisen
equation of state:
\begin{equation}
\label{PMMAeos}
p = \frac{\rho_0c_0^2}{s(1-s\eta)}\Big(\frac{1}{1-s\eta}-1\Big),
\end{equation}
where $\eta=1-\frac{\rho_0}{\rho}$, $c_0$ is the material sound speed, $\rho_0$
is the reference density and $s$ a single experimentally determined coefficient.
For PMMA, these parameters take the values:
\begin{equation}
\label{PMMAeos2}
\rho_0 = 1180\SI{}{\kilogram \per \meter^3}, \qquad c_0 = 2260\SI{}{\meter
\per \second} \qquad \text{and} \qquad s=1.82.
\end{equation}
The low-conductivity material we consider in this work which can be considered
an isotropic approximation to carbon fibre reinforced plastic, follows the
same equation of state as the aluminium, with a lower conductivity value
of $41260\SI{}{\siemens \per \meter}$.

\subsection{Solving each for phase}
As all the systems of equations are in a hyperbolic form, we can
 use finite volume methods to solve for each
phase. This will give solutions at all pure-material cells. Validation
for the individual components can be found in previous work
\cite{michael2018multi,millmore2019,frederik}.  The next step, described in the
section that follows is how to apply material boundary conditions and
fill in the cells adjacent to the material boundary, for which we need
information from both materials.

\section{The multi-material approach}

In this work, we use level set methods to track the material
interfaces given  by the boundaries of the three
PDE-systems (referred to as materials in this
framework) described in the previous section.  Note
  that in this work there is not a distinct interface between plasma and unionised
  gas, and the two materials are modelled using the same equation of
  state.  The change in state of matter happens gradually over a
  diffuse region, with the degree of ionisation being a function of
  temperature and pressure.

Level set methods only give the location of the interface; they do not
affect the evolution of the material components. The behaviour of the
material components at the interface is modelled by the implementation
of dynamic boundary conditions 
  using the Riemann ghost fluid method  with
  appropriately constructed mixed-material Riemann solvers to solve
the interfacial Riemann problems between materials.

Level set methods determine the location of an
  interface through the zero contour of a signed distance function,
  $\phi(x,y)$, typically referred to as the level set function, given
  by $\Phi = {(x,y)|\phi(x,y)=0}$.  The level set function is an
  implicit representation of the interface, we store only discretised,
  values of the function, $\phi_i$, where the sign of the function can
  determine where a material is present within the computational
  domain.  For simulations with three or more materials, each material
  has its own level set function, and is present in the region
  $\phi_i^m > 0$.  The evolution of $\phi(x,y)$, assuming no mass
transfer across the interface, is given by the advection equation:

\begin{equation}
\frac{\partial \phi}{\partial t} + \mathbf{u} \cdot \nabla \phi=0
\end{equation}
where ${\bf u}_i$ is the velocity of the material
  present in cell $i$.

The Riemann ghost fluid method of Sambasivan and
Udaykumar \cite{sambasivan2009ghost1,sambasivan2009ghost2} is
 used to model the behaviour of the material
component at the interface. This method, in contrast to the original
ghost fluid method \cite{fedkiw1999non}, uses the
  solution to a mixed-material problem to compute the dynamic boundary
  conditions at ghost-cell states adjacent to the interface. For
every cell $i$ adjacent to the interface 
the following procedure is used:

\begin{enumerate}
\item locate the interface within the cell at the point $P=i+\phi\nabla \phi$
\item project two probes into the adjacent materials, reaching the points $P_1=P+\mathbf{n}\cdot\Delta x$ and $P_2=P-\mathbf{n}\cdot\Delta x$
\item interpolate states at each point using information from the surrounding cells 
\item solve a mixed Riemann problem (as described in Sec.\ \ref{Sec:MRS}) between the two states to extract the state of the real-material cells, adjacent to the interface (left star state $\mathbf{W^*_L}$, in Fig.\ \ref{Fig:OSRPallaire}) 
\item replace the state in cell $i$ by the computed star state.
\end{enumerate}

After the above procedure is followed for each material, a
fast-marching method is used to fill in the ghost cells for each
material.

\subsection{Mixed Riemann solvers}
\label{Sec:MRS}

In this section we describe how the mixed-material Riemann
problems at material interfaces are
solved (step 4 in the procedure described above). A
  separate mixed-material Riemann problem needs to be constructed for
  each pair of governing equations.  We present mixed-material Riemann
  problem solutions for two key material pairs.  The remaining solvers
  follow these methods closely, and can be found in previous work.
The Riemann solver at the material interface takes two states from the
two different materials as input, these states can be
modelled by different mathematical models. The
  solution to a given problem provides a one-sided solution to the
interface-adjacent (star) state. This solution is
based on the characteristic equations computed from the
mathematical system describing  one of the materials
and by invoking appropriate `boundary conditions' between the two
materials at the interface.   Without loss of
  generality, we consider the characteristic equations to describe the
  material to the left of the interface. In this
  section, we first describe how the Plasma and Gas model is coupled
  with the liquid model (i.e., a simple Euler system) and then with a
  full elastoplastic solid system. The remainder combinations should
  be directly  deducible from these or found in
  \cite{michael2018multi}.

\subsubsection{Plasma-Gas coupled with Liquid}
\label{GasLiquid}

 We consider a material interface, on the left side of
  which is a plasma governed by the equations of Sec.\ \ref{plasma}
  and an our tabular equation of state, and on the right side is a
liquid material governed by the Euler equations and an equation of
state in the form of ideal gas or more complex. Hereforth we assume
that we are currently solving for the plasma material (in GFM
terminology, the `real' material). At the material boundary a Riemann
problem is solved between the left plasma and the right liquid Euler
system to provide the star state for the real material. We use a
Riemann solver that takes into account the two different materials and
all the wave patterns in the Euler system, as described in this
section.

We write the hyperbolic part of the plasma model in primitive form as $\mathbf{W}_t+\mathbf{A(W)W}_x=\mathbf{0}$, with
\begin{equation}
\label{EulerJacobian}
\mathbf{W} = \left[
\begin{array}{c}
\rho  \\
u \\
w \\
p \\
\lambda
\end{array} \right], \quad \mathbf{A}(\mathbf{W}) = \left(
      \begin{array}{ccccccc}
        u & \rho & 0 & 0 & 0\\
        0 & u & 0 & 1/\rho & 0  \\
        0 &  0 & u & 0 & 0 \\
        0 & \rho c^2 & 0 & u & 0 \\
         0 &  0 & 0 & 0 & u \\ 
      \end{array} \right).
  \end{equation}\\
  \\
  The Jacobian matrix $\mathbf{A}(\mathbf{W}) $ has eigenvalues $\mu_1 =\mu_2=\mu_5=u$,
  $\mu_3=u-c$ and $\mu_4=u+c$, where $c$ is the
    soundspeed of the material.
  \\
  \\
  The right eigenvectors are
\begin{equation}
\label{EulerEvecsR}
\mathbf{r}_1 = \left(
1  ,0 , 0 , 0, 0  
 \right)^T, \quad
\mathbf{r}_2 = \left(
0  ,0 , 1 , 0, 0  
 \right)^T,\quad 
 \nonumber \end{equation}
 \begin{equation}
\mathbf{r}_3 = \left(
1 , -c, 0, \rho c^2, 0   \right)^T,\quad
\mathbf{r}_4 = \left(
1 , c, 0, \rho c^2, 0  
 \right)^T, \quad \mathbf{r}_5 = \left(
0  ,0 , 0 , 0,1 
 \right)^T
\end{equation}
and the left eigenvectors are
\begin{equation}
\label{EulerEvecsL}
\mathbf{l}_1 = \left(
-c^2  ,0 , 0 , 1, 0 
 \right),\quad 
\mathbf{l}_2 = \left(
0 , 0 , 1 , 0, 0 
 \right), \quad
\nonumber \end{equation}
 \begin{equation}
\mathbf{l}_3 = \left(
0  , -\rho c , 0 , 1, 0 
 \right),\quad
\mathbf{l}_4= \left(
0  ,\rho c , 0 , 1, 0 
 \right),\quad
\mathbf{l}_5= \left(
0  ,0,0 , 0 , 1 
 \right).
\end{equation}
\vspace{0.001cm}
\begin{figure}[!t]
\center
\includegraphics[width=0.7\textwidth]{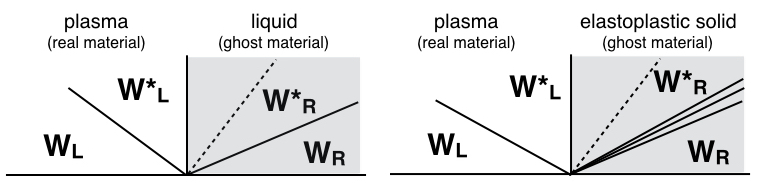}
\caption{The Riemann problem at the material interface between a
  plasma and a fluid (e.g., liquid) or solid elastoplastic
  material. The shaded region represents the initial ghost region\ and
  the white region the initial `real' material region in GFM
  terminology. We are solving for the material in the white, left
  region. For plasma, the one-sided solution 
    comprises two or more degenerate and hence overlapping waves.}
\label{Fig:OSRPallaire}
\end{figure}
Characteristics define directions $\frac{dx}{dt}=\mu_j$, in which
\begin{equation}
\mathbf{l}^{(i)}\cdot d\mathbf{W} = 0, \quad \text{where} \quad 
d\mathbf{W}= \left(d\rho ,  du , dw , dp, d\lambda \right)^T.
\end{equation}
\vspace{0.001cm}
So, along   $\frac{dx}{dt}=\mu_j=u$ for $j=1,2$ and 5  we obtain, respectively:
\begin{eqnarray}
\label{HybridCharacteristics2} 
dp-c^2 d\rho&=&0,\\
\label{HybridCharacteristics2b}
dw&=&0,\\
d\lambda&=&0.
\end{eqnarray}
And along  $\frac{dx}{dt}=\mu_3=u-c$ and $\frac{dx}{dt}=\mu_4=u+c$, we obtain:
\begin{eqnarray}
\label{HybridCharacteristics3}
dp-\rho c\, du&=&0 \\
\label{HybridCharacteristics3b}
\text{and} \quad dp+\rho c\,du&=&0.
\end{eqnarray}

For a single phase described by the ideal gas or stiffened gas equation of state, the characteristic equations can be integrated directly, as the expressions for the sound speed are simple. However, for more complex or tabular equations  this might not be possible. In such case, one can obtain an approximate mixed Riemann solver by replacing the differentials with the difference of the initial and the final state without integrating (i.e.\ across the characteristics), as presented below.
\\
\\
Using $dp +  \rho c\, du=0$, we connect the states $\mathbf{W}^*_L$ and $\mathbf{W}_L$ to obtain:
\begin{equation}
\label{eqLhybrid}
p^*_L-p_L = - \rho_L c_L(u^*_L - u_L).
\end{equation}
Using $dp -  \rho c\, du=0$, we connect the states $\mathbf{W}^*_R$ and $\mathbf{W}_R$ to obtain:
\begin{equation}
\label{eqRhybrid}
p^*_R-p_R = \rho_R c_R(u^*_R - u_R).
\end{equation}
Pressure and velocity don't change across the material interface, hence $p^*_L=p^*_R=p^*$
and $u^*_L=u^*_R=u^*$. Applying these conditions to (\ref{eqLhybrid}) and (\ref{eqRhybrid})
we obtain two expressions for $p^*$ and $u^*$:
\begin{equation}
\label{ustarpstarHybrid}
p^* = p_L - \rho_L c_L(u^*-u_L) \quad \text{and} \quad 
u^* = \frac{p^*-p_R}{\rho_Rc_R} + u_R.
\end{equation}
Solving the above two simultaneously, we obtain an expression for the pressure in the star region:
\begin{equation}
\label{pStarHybrid}
p^* = \frac{C_R p_L - C_L p_R-C_RC_L(u_R-u_L)}{C_R+C_L},
\end{equation}
where $C_L = \rho_Lc_L$ and $C_R=\rho_Rc_R$.

To calculate the left fluid (here gaseous) state, connect states $\mathbf{W}^*_L$ and $\mathbf{W}_L$, using equation (\ref{eqLhybrid}) and $dp-c^2d\rho=0$ to obtain:
\begin{eqnarray}
\label{ustarHybrid}
u^* &=& \frac{p_L-p^*}{\rho_Lc_L} + u_L \quad \text{and}\\
\label{rhostarHybrid}
 \rho^*_L &=& \frac{p^*-p_L}{c_L^2} + \rho_L.
\end{eqnarray}
Using the remaining characteristic equations we obtain \begin{equation}\label{otherStarHybrid}
w^*=w_L  \quad \text{and} \quad \lambda^*=\lambda_L
\end{equation} 
Equations (\ref{pStarHybrid})--(\ref{otherStarHybrid}) give the full
state in the left star region. The values of $C_L,C_R$ and $c^2_L$ in
equations (\ref{pStarHybrid})--(\ref{rhostarHybrid}) are constant
approximations of the $(\;)_L$ and $(\;)_R$ values, although other approximations can be taken.

\subsubsection{Plasma-Gas coupled with Solid}
\label{GasSolid}
\label{Sec:HybridMRSsolid}  We now consider an
  interface on the left of which is again a plasma, governed by the
  equations of Sec.\ \ref{plasma} and an our tabular equation of state
  (plasma-gas material) and on the right side is a material governed
by the elastoplastic solid equations (the solid material) of Sec.\
\ref{solids}.  We develop a Riemann solver that takes into account the
two different materials to determine the star state in the plasma-gas
material.

We follow a similar procedure to that in Sec.\ \ref{GasLiquid}. Referring to Fig.\ \ref{Fig:OSRPallaire}, $\mathbf{W}_L$ corresponds to the original plasma state, $\mathbf{W}_R$ to the original \
elastoplastic state and $\mathbf{W}^*_L$ to the plasma star state that we are looking to compute in this Riemann solver. Since we are solving for the plasma as the real material, the Riemann problem still\
 has three types of waves (two non-linear and two overlapping linear in 1D). The same characteristic relations (\ref{HybridCharacteristics2})--(\ref{HybridCharacteristics3b}) are defined as before and we use
the approach of representing the differentials with the state difference.
 Connecting fluid states  $\mathbf{W}^*_L$ and $\mathbf{W}_L$ using  $dp +  \rho c\, du=0$ and solid states $\mathbf{W}^*_R$ and $\mathbf{W}_R$ using  $dp -  \rho c\, du=0$ we obtain a mixed-material expression for $p^*$:
\begin{equation}                                                                                                                                                                                            
\label{pstarHybridSolid}                                                                                                                                                                                    
p^*=\frac{{u_S-u_F + \frac{1}{\rho_S}(\mathbf{Q}^{-1}\mathcal{D}^{-1}\mathbf{Q})^S_{11}\sigma^S_{11}+\frac{p_F}{\rho_Fc_F}}}{\frac{1}{\rho_Fc_F}-\frac{1}{\rho_S}(\mathbf{Q}^{-1}\mathcal{D}^{-1}\mathbf{Q}\
)^S_{11}\sigma^S_{11}},                                                                                                                                                                                     
\end{equation}
where $\mathbf{Q}$ is an orthogonal matrix and $\mathcal{D}$ is the diagonal matrix of positive eigenvalues for the solid system, such that the elastic acoustic tensor is defined by:
\begin{equation}                                                                                                                                                                                            
\Omega_{ij}=\frac{1}{\rho}\frac{\partial \sigma_{1i}}{\partial F^e_{jk}}{\mathbf{F^e}}_{1k}=\mathbf{Q}^{-1}\mathcal{D}^{-1}\mathbf{Q}.                                                       
\end{equation}
Considering $p_R=\sigma^S_{11},(\mathbf{Q}^{-1}\mathcal{D}^{-1}\mathbf{Q})^S_{11}=1/c_R$ and $C_R = \rho_Rc_R$ and $C_L=\rho_Lc_L$ we obtain equation (\ref{pStarHybrid}).

Then, using $dp+c^2d\rho=0$ to  connect states $\mathbf{W}^*_L$ and $\mathbf{W}_L$ we obtain equations (\ref{ustarHybrid})--(\ref{rhostarHybrid}) and values for $u^*_L$ and $\rho^*_L$, using the remaining characteristic equations we obtain equations (\ref{otherStarHybrid}) and values for $w^*$ and $\lambda^*$. At the interface, we apply conditions
\begin{equation}                                                                                                                                                                                            
\label{slipFluid}                                                                                                                                                                                           
p^*_L = \sigma^*_{R,11}, \quad u^*_L=u^*_R.                                                                                                                                                                 
\end{equation}
To improve accuracy, an iterative approach as described in \cite{michael2018multi} can be used. The coupling of the remainder material combinations follow from the above. Validation for each mixed Riemann solver can be found in previous work \cite{michael2018multi,frederik}.

\section{Lightning strike on elastoplastic substrates }
In this section we exercise our multi-physics methodology applied to
the simultaneous solution of four states of matter to study
lightning-strike scenarios that could lead to the ignition of a
combustible gas, in a metal container containing liquid and gaseous
phases. We consider a closed container made of
aluminium, or a similar metal with lower electrical
  conductivity, and the cases in which the container is coated with
an upper dielectric paint layer.  We can consider
  either this dielectric layer, or the entire top of the container
  (dielectric and metal) to be pre-damaged (opened)}. In the metal
container we consider a  liquid material upon which
  is an upper region of a combustible gas.  All
  simulations are performed in cylindrical symmetry, with the $z$-axis
  the axis of symmetry.

\subsection{Sealed aluminium substrate}

\begin{figure}[!htb]
\centering
\begin{subfigure}[b]{0.4\textwidth}
                \includegraphics[width=\textwidth]{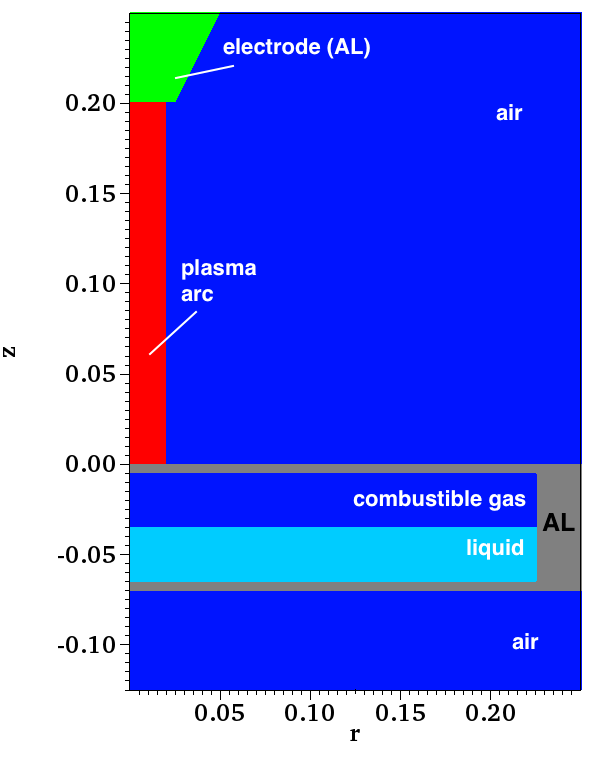}
\caption{Single, sealed, aluminium substrate in grey. \phantom{Single, sealed, aluminium substrate in grey.}}
                \label{IC_ALwoGap}
        \end{subfigure}\hspace{0.5cm}
        \begin{subfigure}[b]{0.4\textwidth}
                \includegraphics[width=\textwidth]{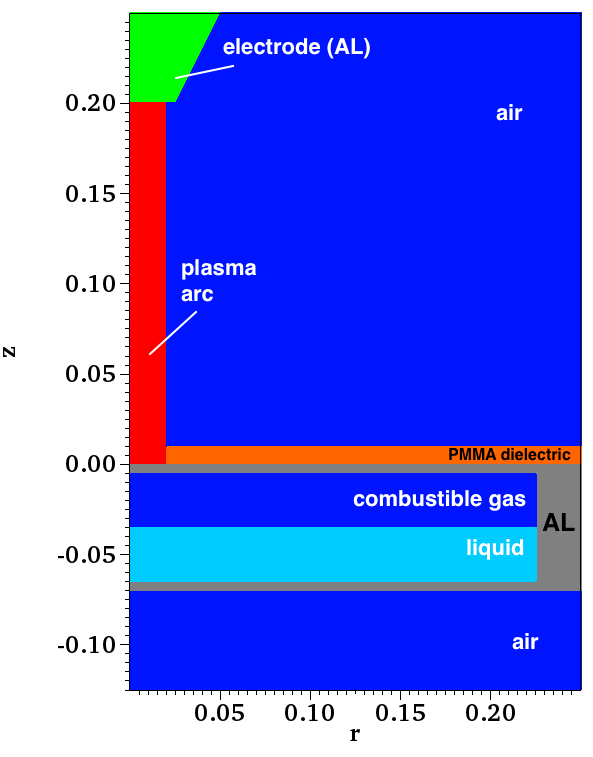}
\caption{ Sealed aluminium substrate in grey and dielectric layer with a gap in orange.}
                \label{IC_ALPMMAwoGap}
        \end{subfigure}
\caption{The setup of the simulations with (right) and without (left) a vented dielectric. All distances are in meters.}
\label{ICmaterialscwoGap}
\end{figure}

In this section we consider the simulation of  plasma
  arc attachment to an undamaged aluminium substrate, underneath
which a layer of a combustible gas and a layer of liquid are
placed. As there is no gap or hole on the  aluminium,
the  plasma arc can only interact indirectly with the
combustible gas; any effects have to propagate through the
 aluminium substrate. Beneath the two fluids lies
another layer of the same  aluminium, forming a
closed metal container, as the aluminium, both on
top and bottom is extended to the right end of the domain. A schematic
of the initial configuration in 2D is shown in Fig.\ \ref{IC_ALwoGap}.

\begin{figure}[!htb]
\centering
\begin{subfigure}[b]{0.4\textwidth}
                \includegraphics[width=\textwidth]{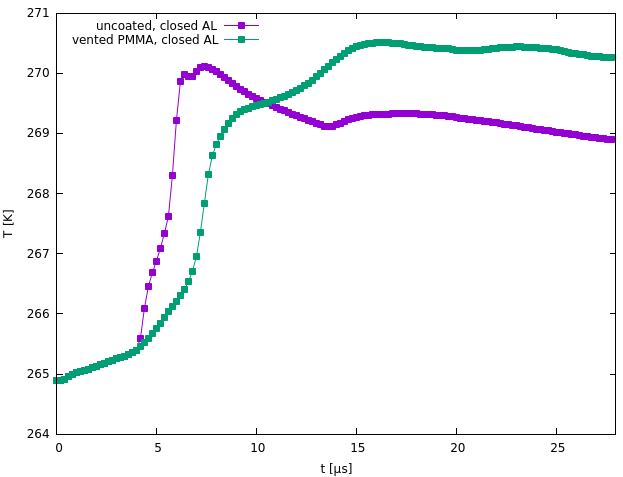}
\caption{Maximum temperature in the combustible gas over time.}
                \label{maxT}
        \end{subfigure}\hspace{0.5cm}
        \begin{subfigure}[b]{0.4\textwidth}
                \includegraphics[width=\textwidth]{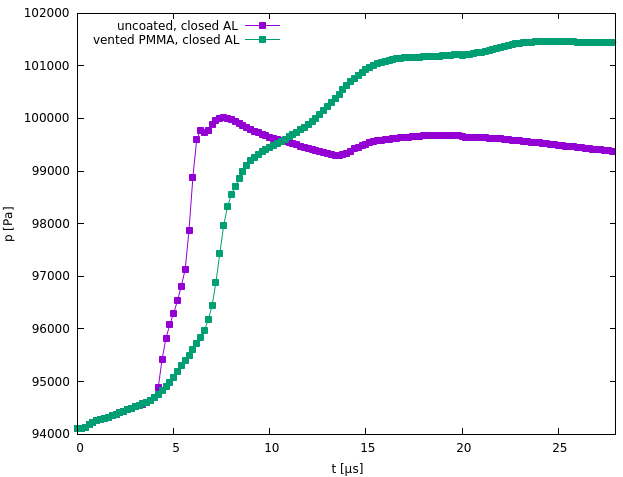}
\caption{Maximum pressure in the combustible gas over time.}
                \label{maxP}
        \end{subfigure}
\caption{Maximum (a) temperature and (b) pressure in the combustible gas throughout the simulation, for an uncoated {\color{black} aluminium} substrate (purple) and PMMA-coated {\color{black} aluminium} substrate, with a gap on the PMMA dielectric (green).}
\label{maxTP}
\end{figure}

\begin{figure}[!htb]
\centering
\begin{subfigure}[b]{0.4\textwidth}
                \includegraphics[width=\textwidth]{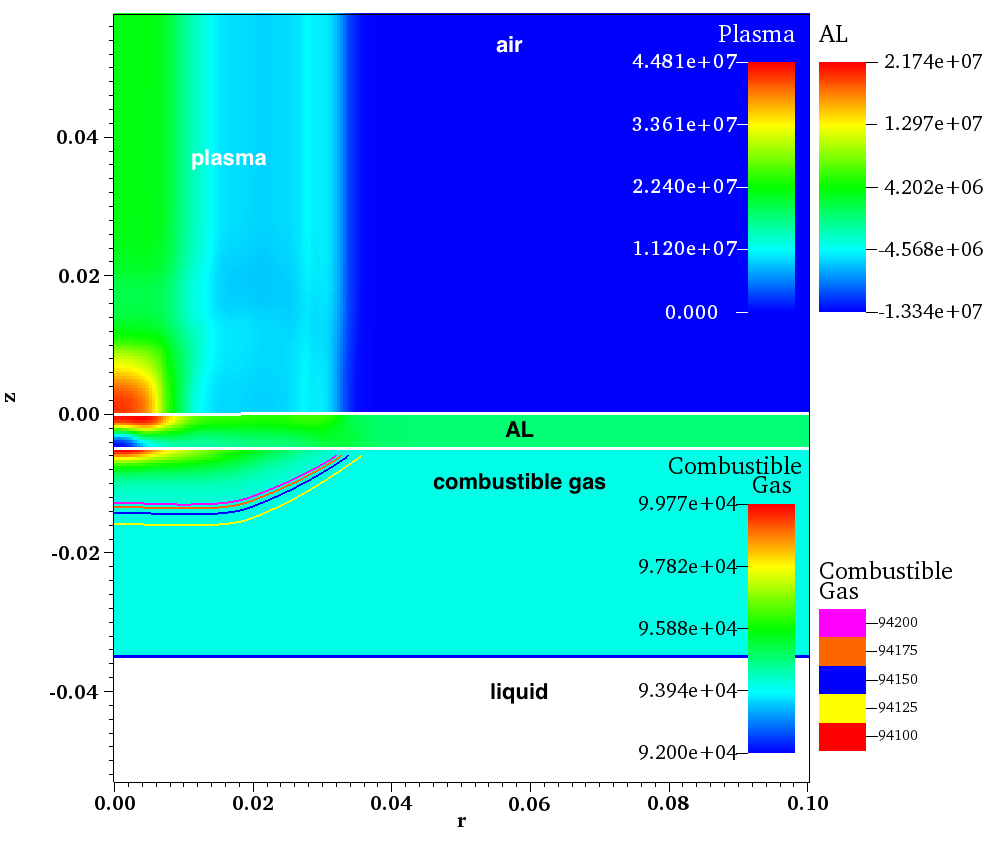}
\caption{$t=\SI{6.4}{\micro \second}$ }
                \label{}
        \end{subfigure}\hspace{0.5cm}
        \begin{subfigure}[b]{0.4\textwidth}
                \includegraphics[width=\textwidth]{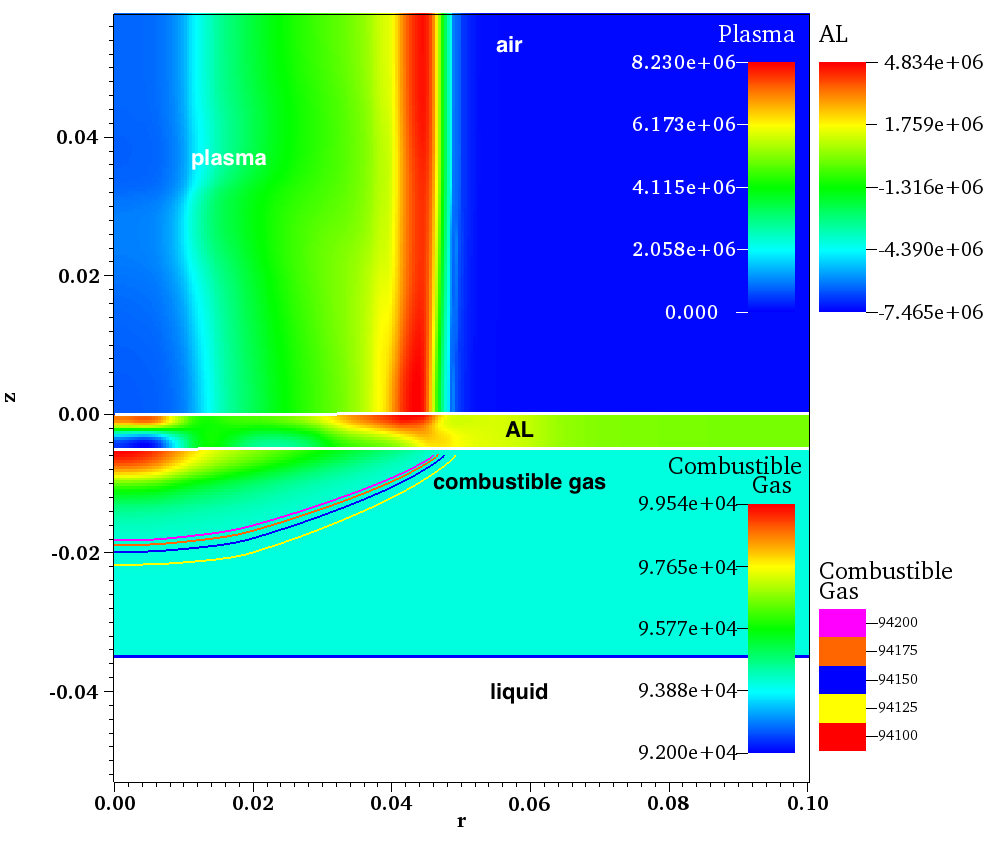}
\caption{$t=\SI{11}{\micro \second}$}
                \label{}
        \end{subfigure}
\caption{Pressure distribution in plasma, aluminium and combustible
  gas for the uncoated, sealed {\color{black} aluminium case}. The material boundary of the aluminium substrate is indicated by the white line, while the interface between the combustible gas and the liquid (whose pressure is not shown here) is indicated by the blue line. }
\label{2Dpressure_closedAL}
\end{figure}

For the gas  within the container to start reacting,
the temperature needs to rise sufficiently to activate the
temperature-dependent Arrhenius term in (\ref{arrhenius}). In this
case, even though the gas is sensitive ($T_A=\SI{8000}{\kelvin}$), the
initiation of reaction requires temperatures of the order of
$1000\SI{}{\kelvin}$.   
  Over the duration of this
simulation the temperature increases only a few degrees, up to
$\SI{270}{\kelvin}$, as seen in Fig.\ \ref{maxT} (purple). The
temperature rise in this simulation comes from the equation of state
responding to the pressure rise of the material, hence the two
quantities follow the same trend. The pressure rise in the combustible
gas is due to a pressure wave propagated from the plasma arc
 through the  aluminium substrate
and then to the gas. In Figure \ref{2Dpressure_closedAL} we study the
pressure waves generated in the early stages of the simulation. The
 rapid growth of the plasma arc induces a pressure
wave in the aluminium of the order of $\SI{2.6}{\mega \pascal}$. The
current  input into the system reaches its peak at
$\SI{6}{\micro \second}$,  at which point we see
maximum pressures of the order of $\SI{85}{\mega \pascal}$ in the
aluminium. The induced pressure in the combustible gas is considerably
lower than that ($\SI{0.1}{\mega \pascal}$), as demonstrated by the
pressure contours and pressure field in Fig.\
\ref{2Dpressure_closedAL}.  This is due to the initial
  shock wave travelling through the aluminium reflecting off of the
  bottom of the layer, generating a strong rarefaction which serves to
  mitigate the pressure increases in the combustible gas.  The
increase of the pressure  over the duration of the
simulation is shown in Fig.\ \ref{maxP} (purple). 
  Since the maximum current  input occurs at
$\SI{6}{\micro \second}$ and thereafter the current profile follows an
exponential decay, we don't expect the pressure and temperature rise
to ignite the material in this scenario. Indeed the $\lambda$-field
remains at the value 1 (i.e., reactants have not been depleted)
throughout the simulation, demonstrating that ignition does not occur
in this configuration, over these timescales.   The
  high electrical conductivity of aluminium serves to mitigate
  temperature rise within the substrate, and therefore even over
  longer timescales, there will not be sufficient temperature
  diffusion to ignite the combustible gas.

\subsubsection{Sealed aluminium substrate, coated with opened dielectric layer}
\begin{figure}[!htb]
\centering
\begin{subfigure}[b]{0.4\textwidth}
                \includegraphics[width=\textwidth]{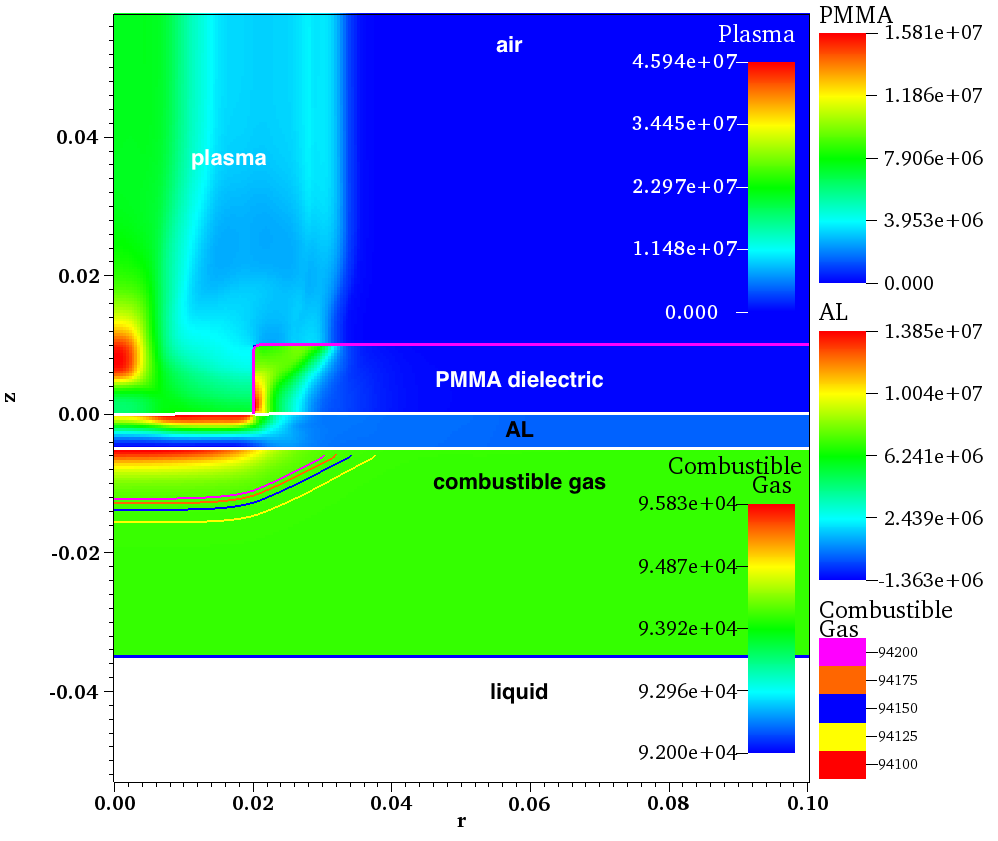}
\caption{$t=\SI{6.4}{\micro \second}$ }
                \label{}
        \end{subfigure}\hspace{0.5cm}
        \begin{subfigure}[b]{0.4\textwidth}
                \includegraphics[width=\textwidth]{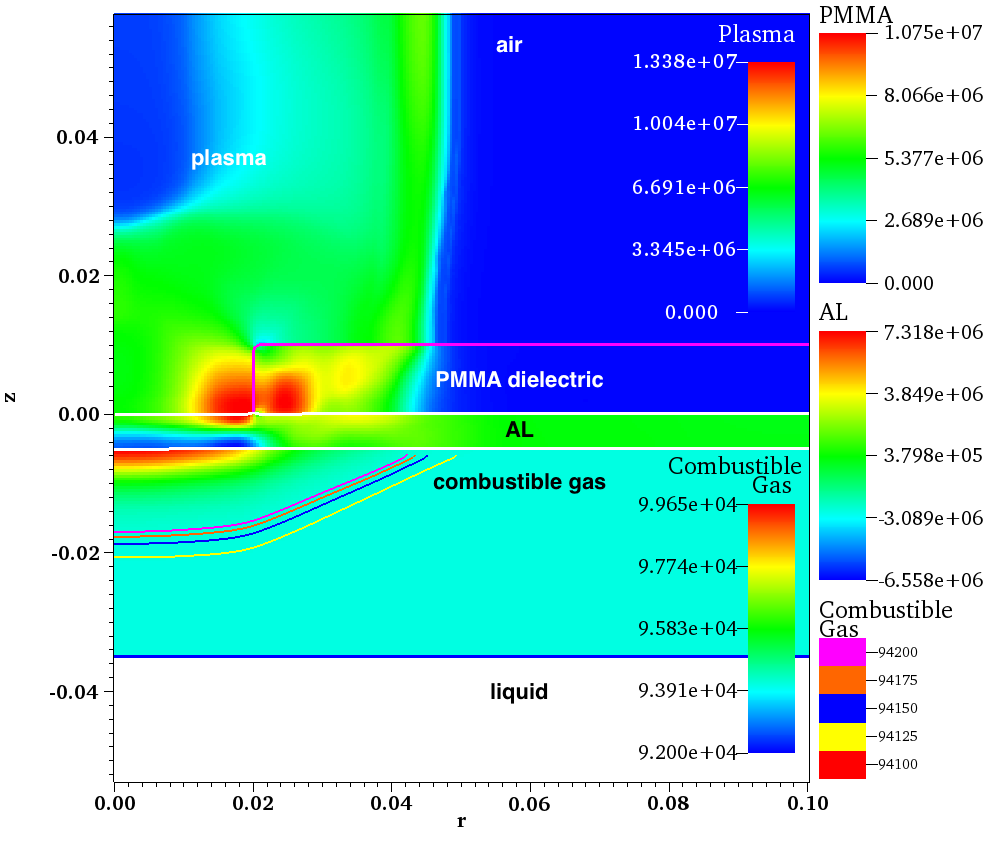}
\caption{$t=\SI{11}{\micro \second}$} 
                \label{}
        \end{subfigure}
\caption{Pressure distribution in plasma, aluminium and combustible gas. The material boundary of the aluminium substrate is indicated by the white line, while the interface between the combustible gas and the liquid (whose pressure is not shown here) is indicated by the blue line and the material boundary of the dielectric is indicated by the fuschia line. }
\label{2Dpressure_closedALventedPMMA}
\end{figure}

We place a dielectric layer, with a thickness of $10\SI[load=text]{}{\milli \metre}$ and a $2\SI[load=text]{}{\milli \metre}$-radius gap at the
centre over the aluminium
substrate,  representative of the situation in which
  there is existing damage to the coated layer (e.g., due to the
  electric breakdown which generates the arc), but not to the metal
substrate, as seen in Fig.\ \ref{IC_ALPMMAwoGap}. We
  use this test to study the pressure and temperature rise in the
combustible gas, as before, to asses the possibility of ignition in
this configuration  over short timescales. The
channel produced by the opening of the dielectric layer can lead to a
funnelling effect for the plasma  which can increase
the pressure locally,  and this is known to increase
 local damage to the metal
substrate.  Therefore we investigate if this effect
is enough to raise the temperatures and pressures in the combustible
material and lead to the ignition of the combustible material. As in
the uncoated  aluminium case, in the duration of this
simulation the temperature increases only a few degrees, up to
$\SI{270}{\kelvin}$, as seen in Fig.\ \ref{maxT} (green). Again, the
temperature rise in this simulation comes from the equation of state
responding to the pressure rise of the material. The pressure rise in
the combustible gas is due to a pressure wave propagated from the
 arc to the  aluminium substrate and then to the
gas. However, the presence of the dielectric restricts the attachment area of the arc, since little current can
  pass through the dielectric layer, and this serves to focus the
  Joule heating effect within the aluminium.  In Figure
\ref{2Dpressure_closedALventedPMMA} we study the pressure waves
generated in the early stages of the simulation. The 
  growth of the plasma arc induces a pressure wave in the aluminium
of the order of $\SI{2.6}{\mega \pascal}$. The maximum pressure in the
aluminium in this case is lower than in the 
  aluminium-only case, reaching values of the order of
$\SI{80}{\mega \pascal}$.   This lower pressure is a
  result of the mechanical interaction between the plasma arc and the
  dielectric layer.  This layer prevents the surface of the aluminium
  from being in contact with the leading shock wave of the arc, thus
  reducing the pressure loading.  However, the concentrated current
  density attachment will lead to higher energy input in the case of a
  coated aluminium substrate.  The induced pressure in the
combustible gas is again considerably lower than that
($\SI{0.1}{\mega \pascal}$), as demonstrated by the pressure contours
and pressure field in Fig.\ \ref{2Dpressure_closedALventedPMMA}. The
maximum pressure in the duration of the simulation is shown in Fig.\
\ref{maxP}. As the maximum current occurs at $\SI{6}{\micro \second}$
and thereafter the current profile demonstrates an exponential decay,
we don't expect the pressure and temperature rise to ignite the
material in this scenario. Again, the $\lambda$-field remains at the
value 1 throughout the simulation, demonstrating that ignition does
not occur in this configuration, over these timescales.

\subsection{Opened Aluminium substrate}

\begin{figure}[!htb]
\centering
\begin{subfigure}[b]{0.4\textwidth}
                \includegraphics[width=\textwidth]{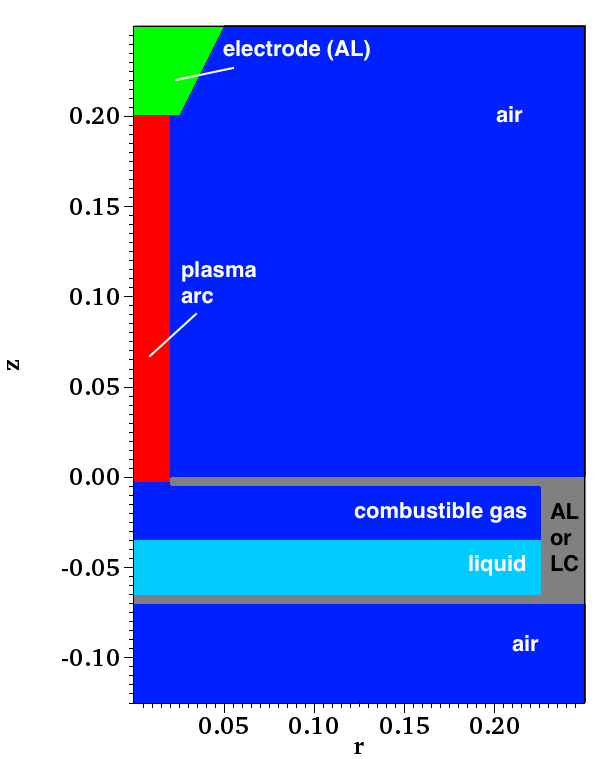}
\caption{Single, uncoated, opened metal (aluminium/low-conductivity) substrate. }
                \label{IC_AL}
        \end{subfigure}\hspace{0.5cm}
        \begin{subfigure}[b]{0.4\textwidth}
                \includegraphics[width=\textwidth]{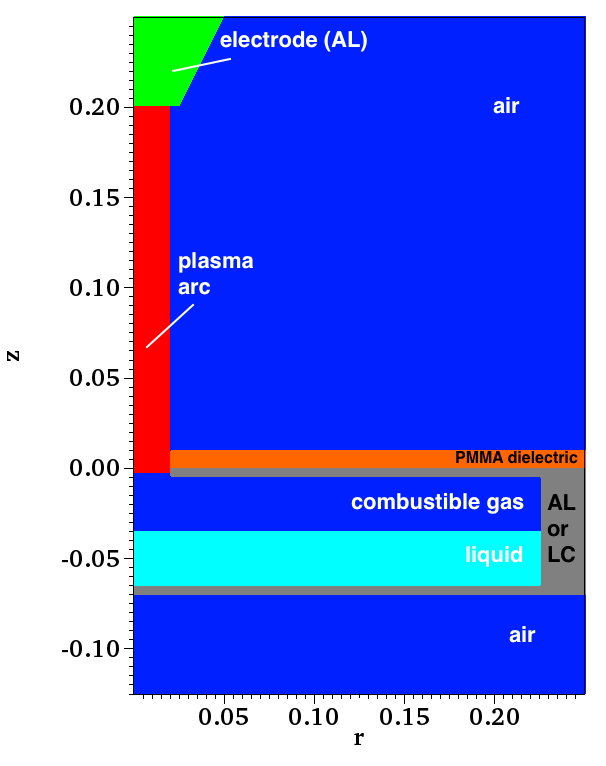}
\caption{Opened metal substrate coated with a dielectric layer.}
                \label{IC_ALPMMA}
        \end{subfigure}
\caption{The setup of the simulations with (right) and without (left)
  a dielectric {\color{black} coating}. All distances are in meters.}
\label{ICmaterials}
\end{figure}

In this section we consider the simulation of  an arc
  attachment to an aluminium substrate that is assumed to have
already been punctured,  beneath which a layer of a
combustible gas and a layer of liquid are 
  present. {\color{black}Since a gap is considered to already exist, either due to an improper lightning protection scheme allowing damage from a previous lightning strike, or by other means, the arc interacts directly with} the combustible gas.  The lower edge of the container is assumed not
  to be damaged, and the aluminium, both on top and bottom is
extended to the right end of the domain. A schematic of the initial
configuration in 2D is shown in Fig.\ \ref{IC_AL}.

\begin{figure}[!thb]
\begin{minipage}{\columnwidth}
        \centering
        \begin{subfigure}[b]{0.5\textwidth}
                \includegraphics[width=\textwidth]{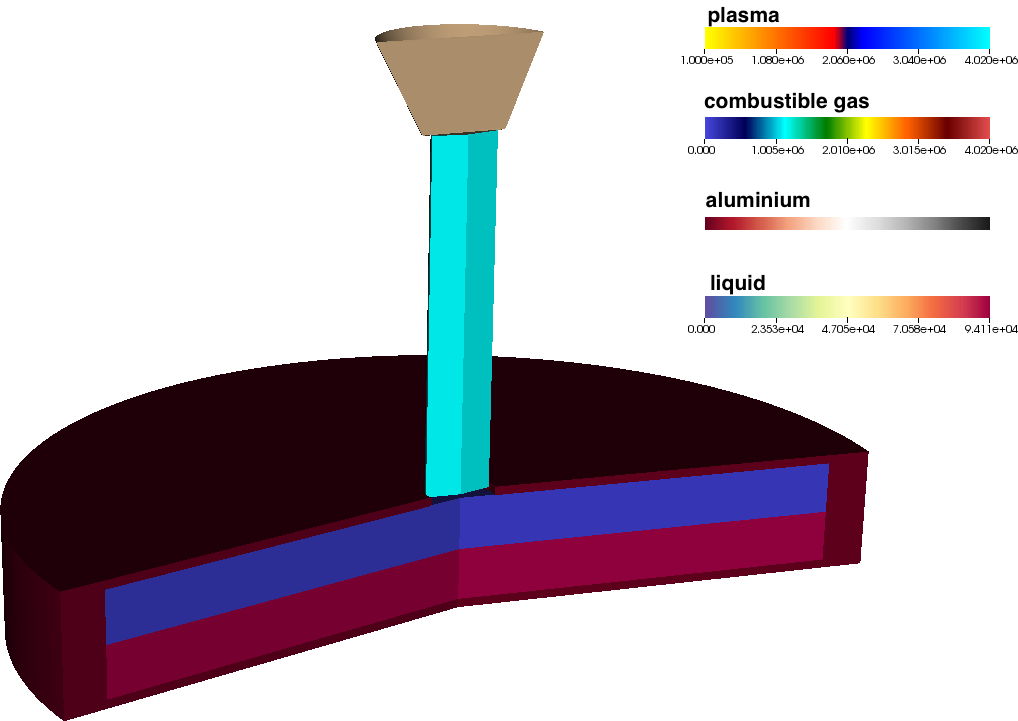}
                \caption{$t=0\SI{}{\micro \second}$}
                \label{AL1}
        \end{subfigure}%
        \begin{subfigure}[b]{0.5\textwidth}
                \includegraphics[width=\textwidth]{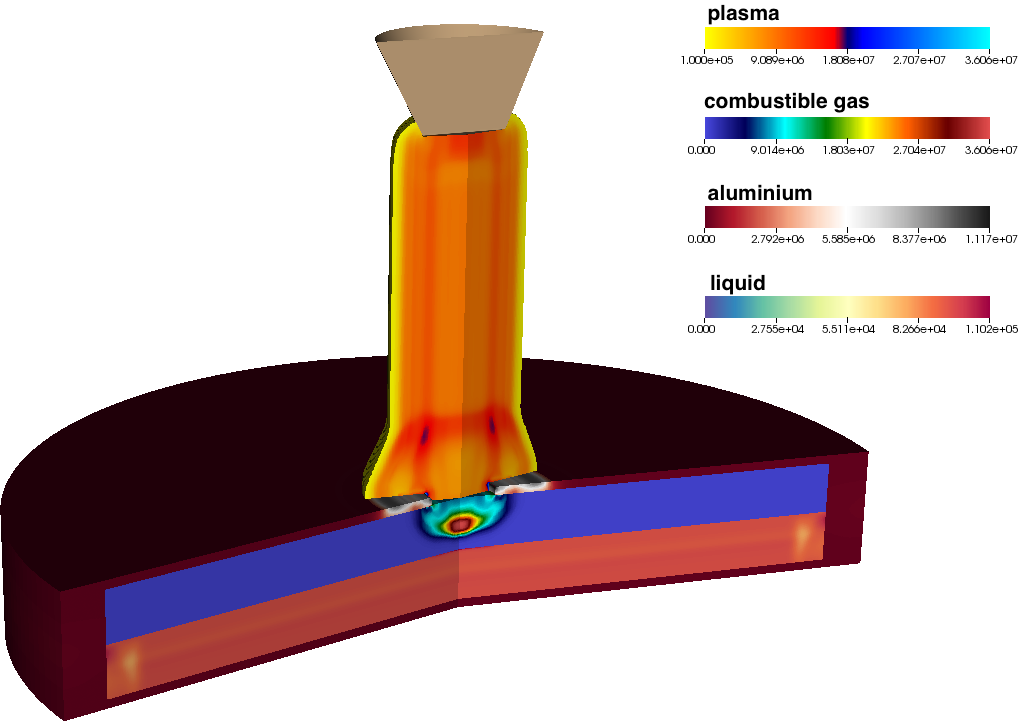}
                \caption{$t=\SI{7.2}{\micro \second}$}
                \label{AL2}
        \end{subfigure}
\end{minipage}
\begin{minipage}{\columnwidth}
        \centering
        \begin{subfigure}[b]{0.5\textwidth}
                \includegraphics[width=\textwidth]{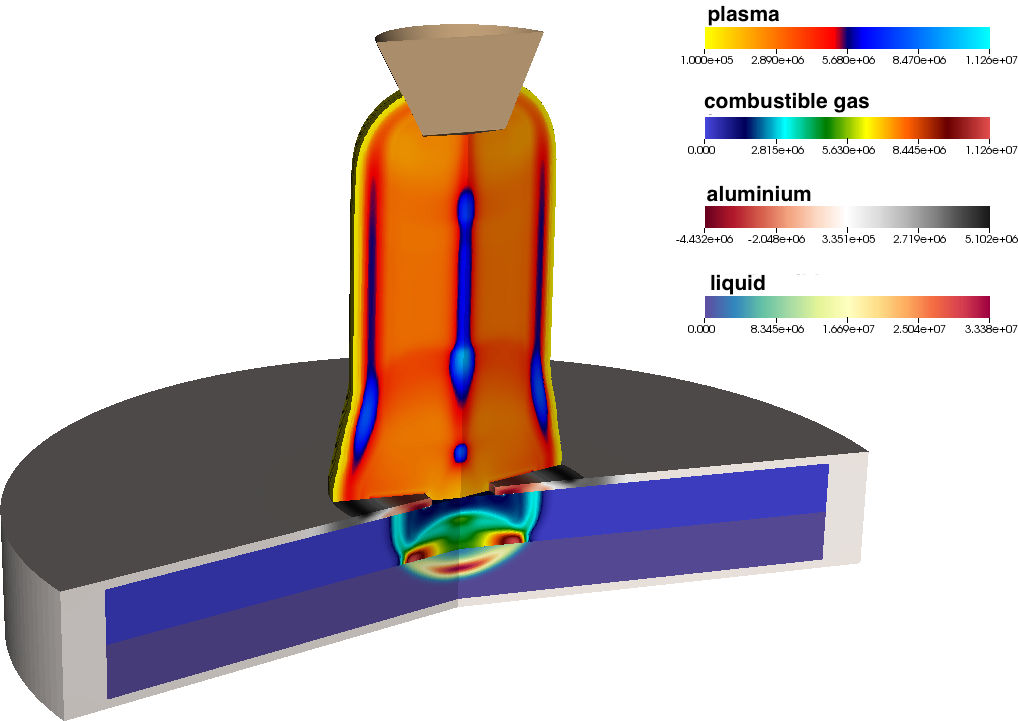}
                \caption{$t=\SI{14}{\micro \second}$}
                \label{AL3}
        \end{subfigure}%
        \begin{subfigure}[b]{0.5\textwidth}
                \includegraphics[width=\textwidth]{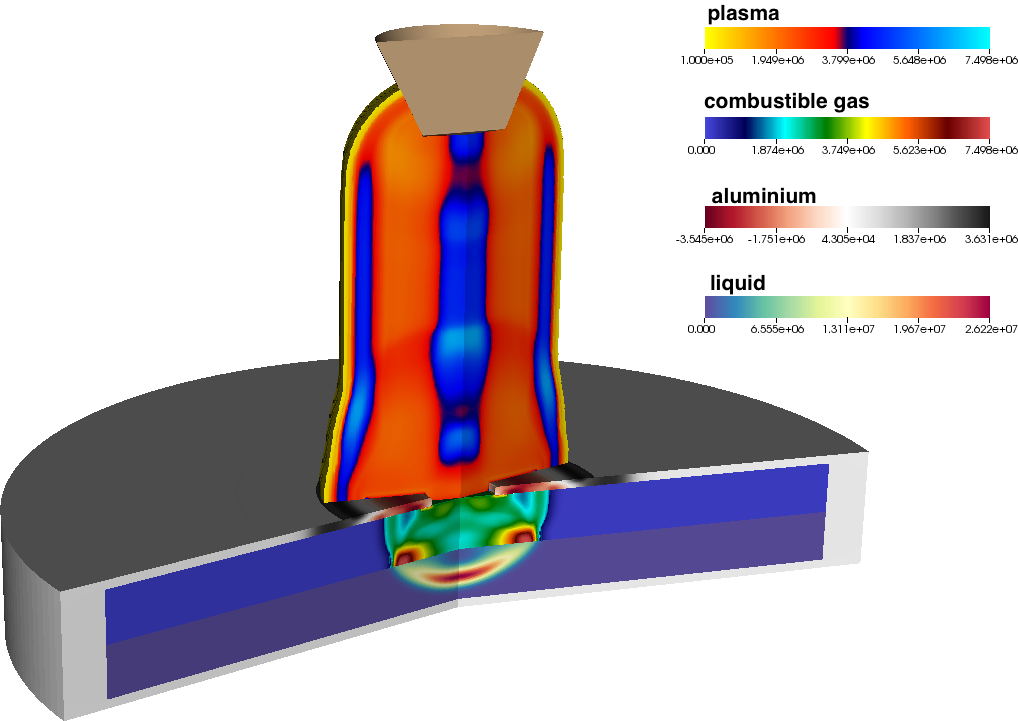}
                 \caption{$t=\SI{16}{\micro \second}$}
                \label{AL4}
        \end{subfigure}
\end{minipage}
\begin{minipage}{\columnwidth}
        \centering
        \begin{subfigure}[b]{0.5\textwidth}
               \includegraphics[width=\textwidth]{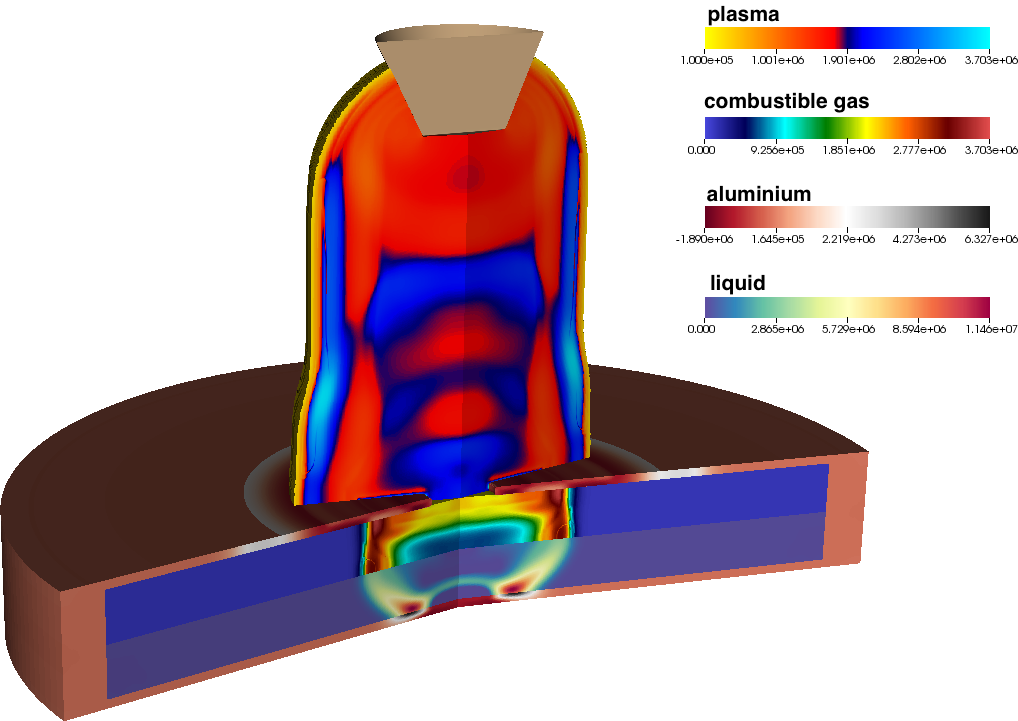}
               \caption{$t=\SI{24}{\micro \second}$}
                \label{AL6}
        \end{subfigure}%
        \begin{subfigure}[b]{0.5\textwidth}
                \includegraphics[width=\textwidth]{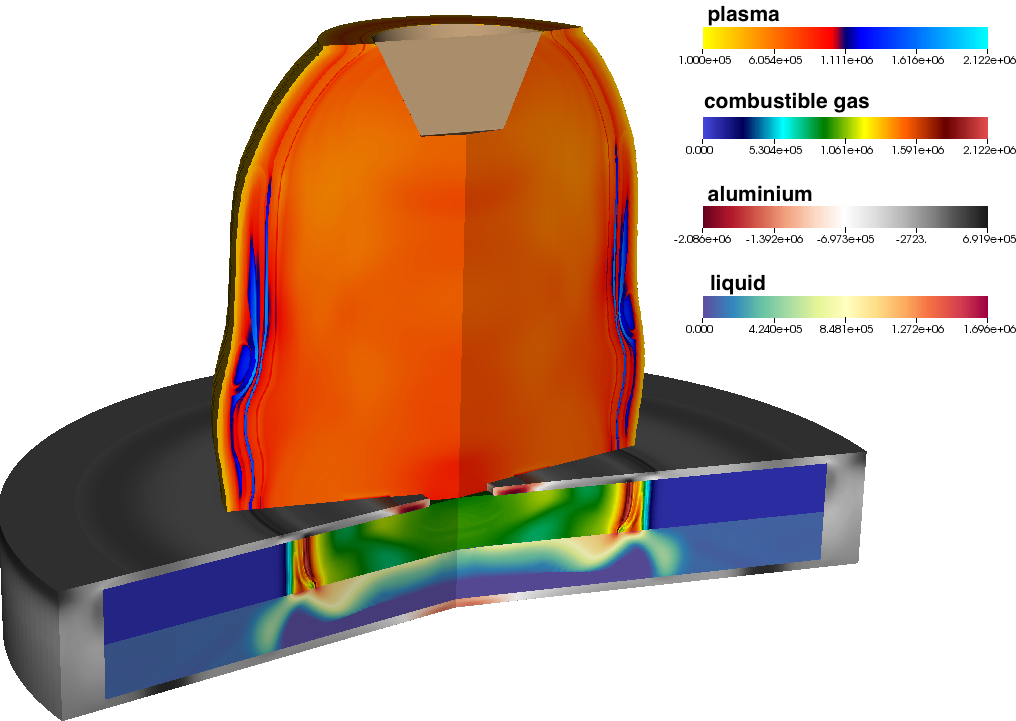}
                \caption{$t=\SI{51}{\micro \second}$}
                \label{AL8}
        \end{subfigure}
\end{minipage}
\caption{Pressure distribution in the plasma, combustible gas, aluminium substrate and liquid, at different times, for the uncoated, opened  aluminium case.} 
\label{ALa} 
\end{figure}

Fig.\ \ref{ALa} shows the evolution of pressure fields in the four
materials. It should be noted that the  plasma arc
  initially attaches to the open-end of the aluminium as seen in
Fig.\ \ref{ALcurrentdensity}. Fig.\ \ref{AL1} shows the initial
conditions for the simulation and in Fig.\ \ref{AL2}
a pressure loading is evident on the aluminium substrate; a shock wave
is seen to propagate in the aluminium. The shock wave 
  progressing through the combustible gas is also shown,
 which raises the pressure by two orders of magnitude
(Fig.\ \ref{AL2}).  
  The shock wave in the combustible gas
reaches the liquid boundary, and as it travels from a low impedance to
a high impedance medium, it generates a  further
shock wave  in the liquid and another
 reflects into the combustible gas (Fig.\
\ref{AL3}).  As this wave propagates upwards
 through the combustible gas,  it
reaches the top of the container where it interacts with the
aluminium, producing another shock in the aluminium and a rarefaction
from the corners of the hole in the aluminium back
into the gas (Fig.\ \ref{AL4}). When the shock wave in the liquid
reaches the bottom of the aluminium container, it 
  also generates a shock in the aluminium and a rarefaction
 back into the liquid (Figs.\ \ref{AL6}). The complex
wave and material interactions continue to late stages (Fig.\
\ref{AL8}).

As the arc evolves over time, the highest pressure remains
at the centre, due to the Joule heating effects,
with the majority of the pressure loading on the combustible gas
occurring
here. A corresponding wave moves through the substrate as the shock wave
imparts a loading effect, though this is substantially lower than at
the centre of the arc. We note that whilst the pressure within the
plasma arc must strictly stay positive, within the substrate, negative
values are experienced. This is because pressure is a component of the
stress tensor, and a solid material can sustain tension, as a result
of rarefaction waves.

\begin{figure}[!t]
\begin{minipage}{0.9\columnwidth}
        \centering
        \begin{subfigure}[b]{0.5\textwidth}
                \includegraphics[width=\textwidth]{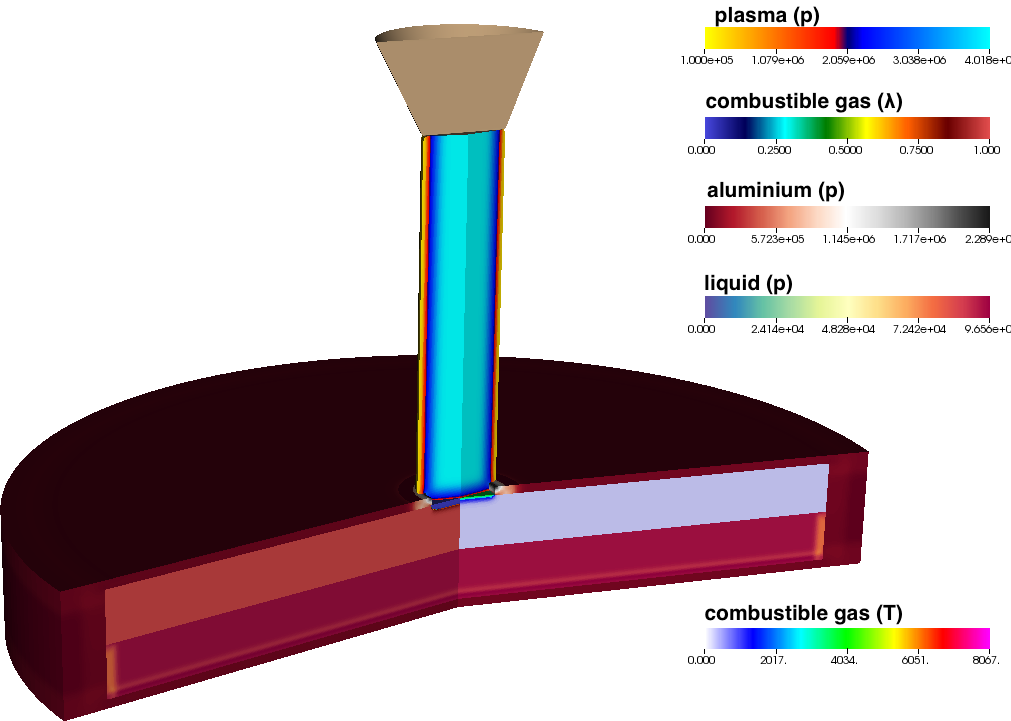}
                \caption{$t=\SI{2}{\micro \second}$}
                \label{AL_TLam1}
        \end{subfigure}%
        \begin{subfigure}[b]{0.5\textwidth}
                \includegraphics[width=\textwidth]{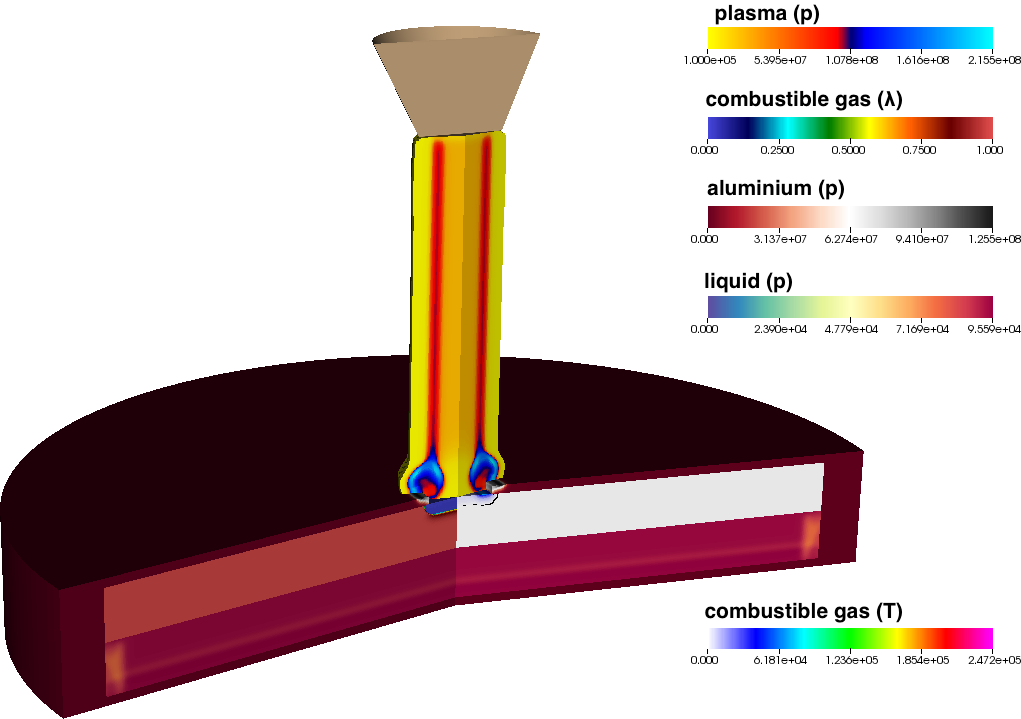}
                \caption{$t=\SI{4}{\micro \second}$}
                \label{AL_TLam2}
        \end{subfigure}
\end{minipage}
\begin{minipage}{\columnwidth}
        \centering
        \begin{subfigure}[b]{0.5\textwidth}
              \includegraphics[width=\textwidth]{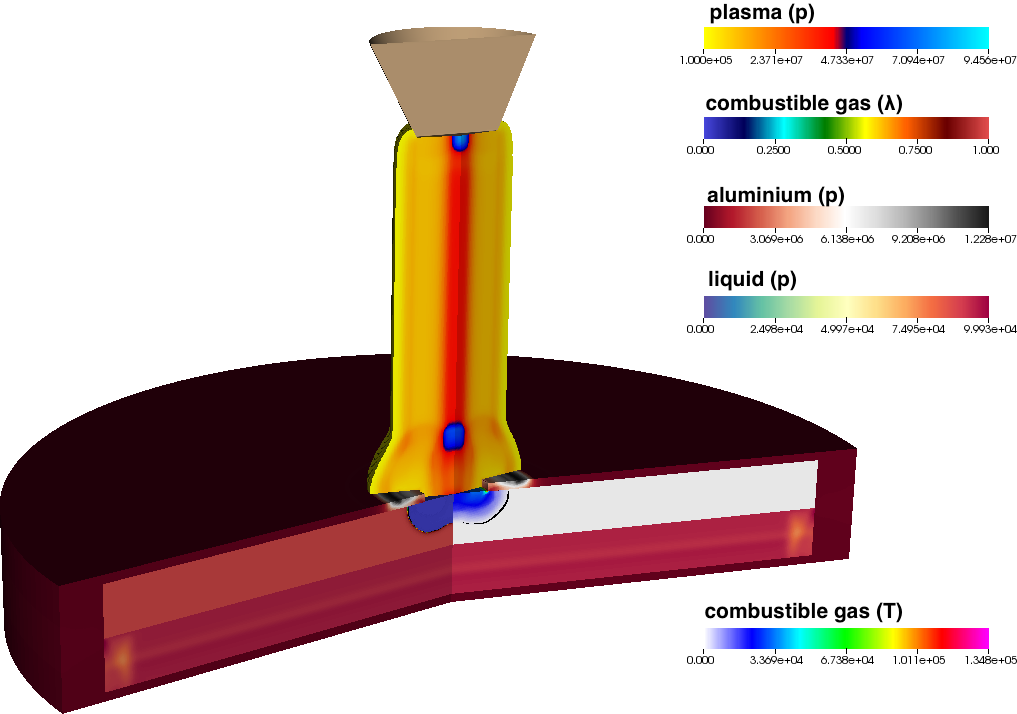}
              \caption{$t=\SI{6}{\micro \second}$}
               \label{AL_TLam4}
        \end{subfigure}%
        \begin{subfigure}[b]{0.5\textwidth}
                \includegraphics[width=\textwidth]{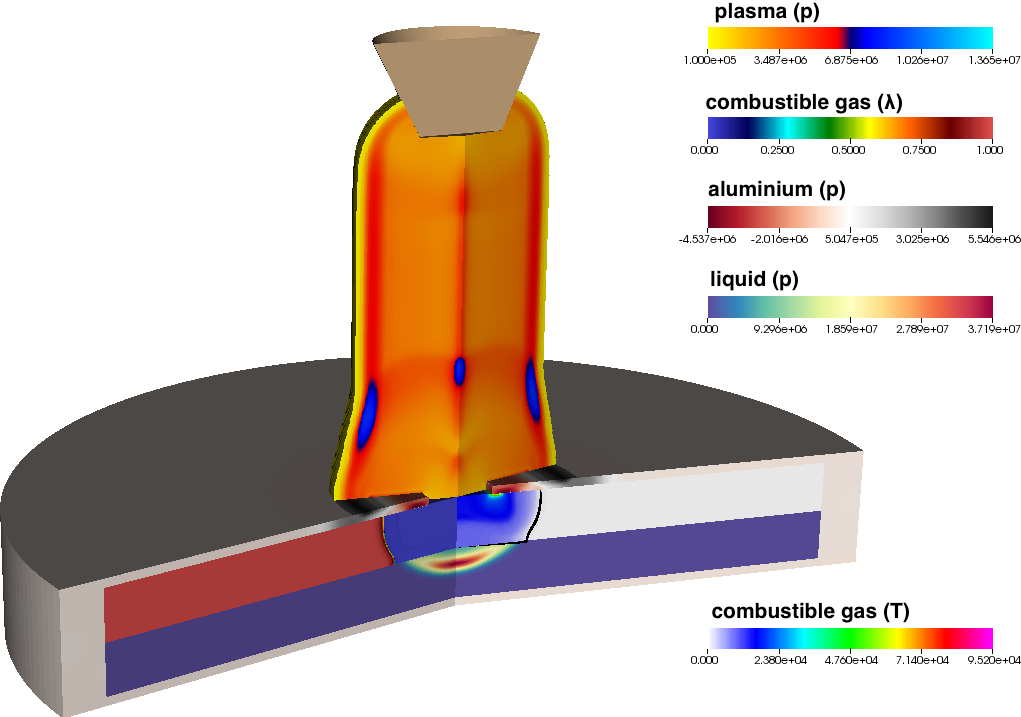}
                \caption{$t=\SI{13.2}{\micro \second}$}
                \label{AL_TLam6}
        \end{subfigure}
\end{minipage}
\begin{minipage}{\columnwidth}
        \centering
        \begin{subfigure}[b]{0.5\textwidth}
                \includegraphics[width=\textwidth]{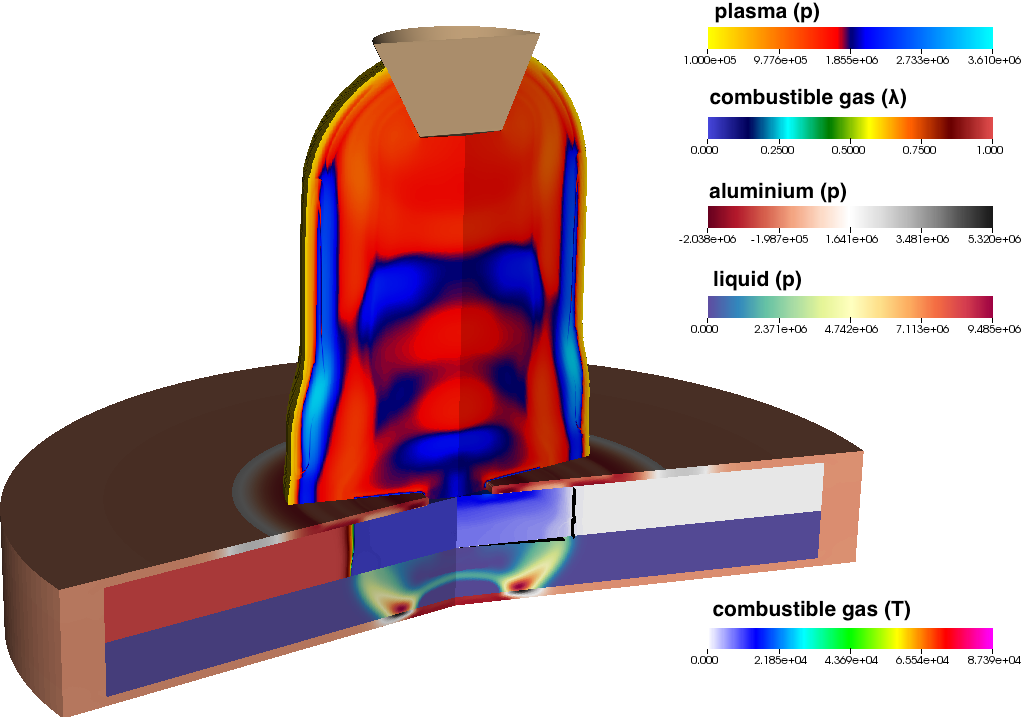}
                \caption{$t=\SI{25}{\micro \second}$}
               \label{AL_TLam8}
        \end{subfigure}%
        \begin{subfigure}[b]{0.5\textwidth}
                \includegraphics[width=\textwidth]{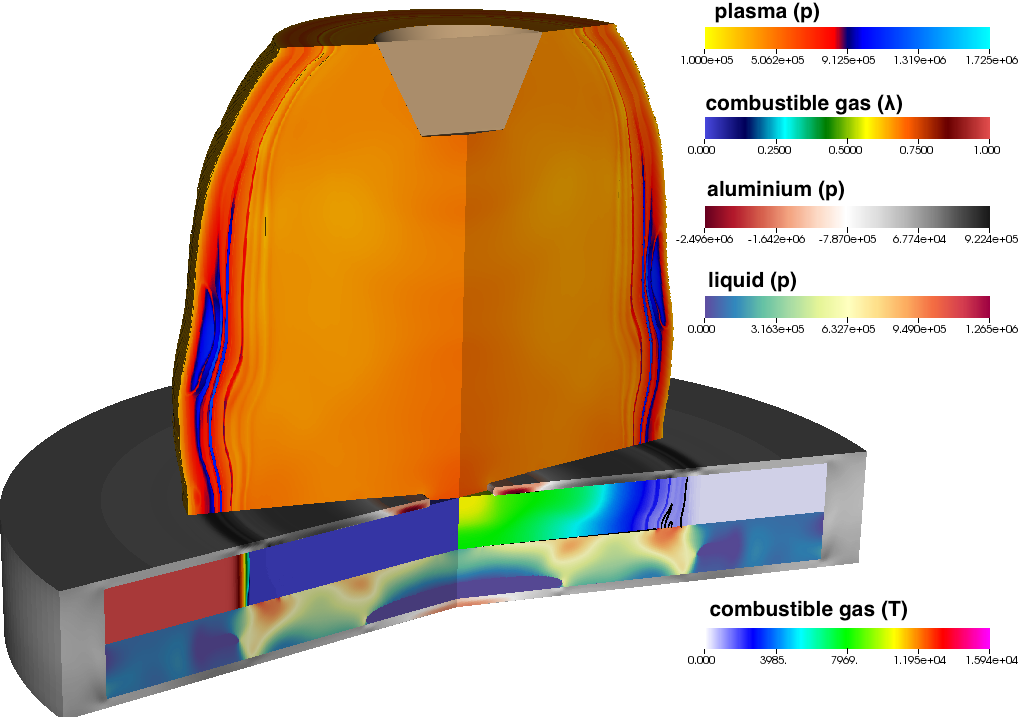}
                \caption{$t=\SI{71}{\micro \second}$}
               \label{AL_TLam9}
        \end{subfigure}
\end{minipage}
\caption{Pressure distribution in the plasma, 
    aluminium substrate and liquid, at different times, for the
  uncoated, opened aluminium case. In the
  combustible gas, the temperature field is shown on the right half
  and the $\lambda$-field on the left half  of each
    plot. On the right half, the contour of temperature of
  1000$\SI{}{\kelvin}$ is also shown. }
\label{ALc} 
\end{figure}

As we are interested in the ignition of the combustible gas, which is
a temperature-driven effect, we present the temperature profiles
(right half) and reaction progress (left half) in Fig.\ \ref{ALc},
 whilst again showing the pressure profile in the
plasma arc, the liquid and the aluminium. Along with the distribution
of temperature in the combustible gas, we present the temperature
contour of $1000\SI{}{\kelvin}$, as the reaction starts at this lower
temperature and grows rapidly as the temperature increases
further. The initial pressure loading is enough to immediately ignite
the material, reaching temperatures of $8000\SI{}{\kelvin}$. 
As the combustible gas is sensitive, this initial pressure
loading leads to its direct combustion.  Therefore,
the high temperature regions on the right of the figures (e.g., Fig.\
\ref{AL_TLam2})  are accompanied by a region of burnt
material shown in the reaction progress variable on the left of the
figure. The shape of the reaction region follows the curved path of
the pressure wave and induced temperature field due to
 the interaction of this wave around the corner of the
  gap in the aluminium. As the reaction front reaches the liquid
interface, it is forced to propagate laterally, thus
giving the curved reaction front seen in Fig.\ \ref{AL_TLam6}.

The rarefaction waves generated at the gas/solid interface work to lower the pressure and weaken the reaction, while the shocks at the liquid/gas interface work to elevate the pressure and strengthen the reaction. Moreover, the dynamic profile of the current density also works to elevate the pressures at the left (back) end of the reaction, continuously altering the rear boundary condition of the region. The reaction continues (Fig.\ \ref{AL_TLam9}) throughout this complex wave interaction until the combustible gas is depleted.

\subsection{Opened, low-conductivity substrate }

 Lightning strikes input a large amount of energy
into the substrate through the Joule heating effect, the
spatial extents over which this occurs is governed
by the electrical conductivity of the material. High-conductivity
materials like aluminium spread this energy over the entire strike
path through the substrate, and thus minimise damage, but the lower
conductivity of composite materials, such as carbon
fibre reinforced plastics (CFRP), means much more
  energy is deposited local to the impact site. Coupled with the
lower thermal conductivity of these materials, this
  leads to substantial heating over diffusion timescales, and hence
lightning strike has the potential to be much more damaging. As a
result, we want to investigate the effect of replacing aluminium with
CFRP on the ignition and detonation of the
combustible gas lying beneath the elastoplastic substrate.
To fully model a CFRP substrate, we need an isotropic
  equation of state corresponding to the orientation of the fibre
  weave comprising the substrate.  In order to model the effects
  within cylindrical symmetry, we consider an isotropic approximation
  to CFRP, in which we consider an electrical conductivity
  corresponding to the direction along the fibre weave of
  $\SI{41260}{\siemens \per \meter}$.  This is substantially lower
  than the $\SI{3e7}{\siemens \per \meter}$ of aluminium.

\begin{figure}[!htb]
\centering
        \begin{subfigure}[b]{0.4\textwidth}
                \includegraphics[width=\textwidth]{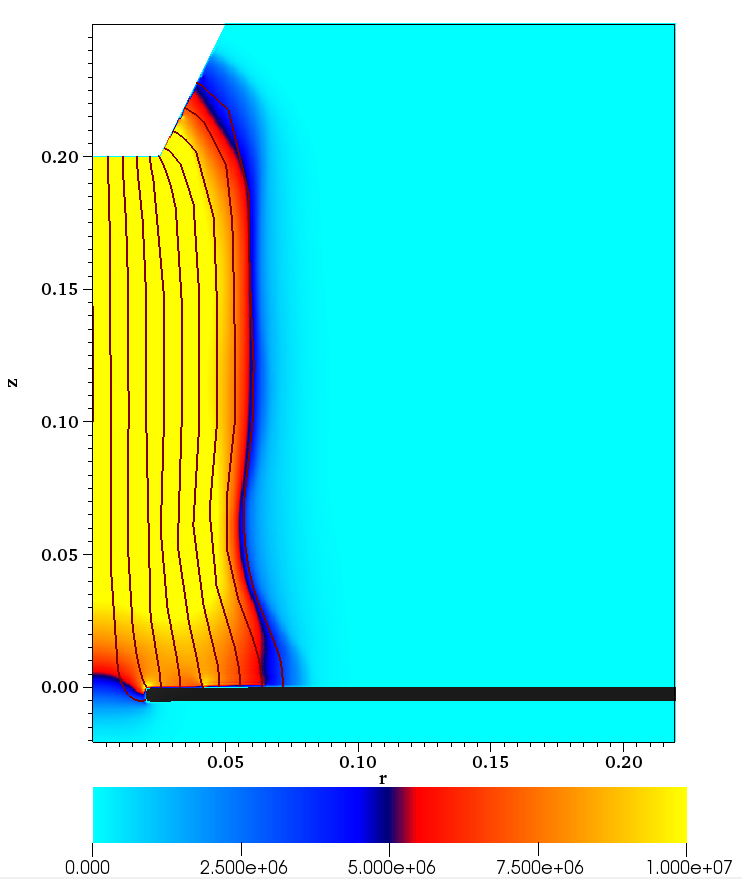}
                \caption{aluminium}
                \label{ALcurrentdensity}
        \end{subfigure}%
        \begin{subfigure}[b]{0.4\textwidth}
                \includegraphics[width=\textwidth]{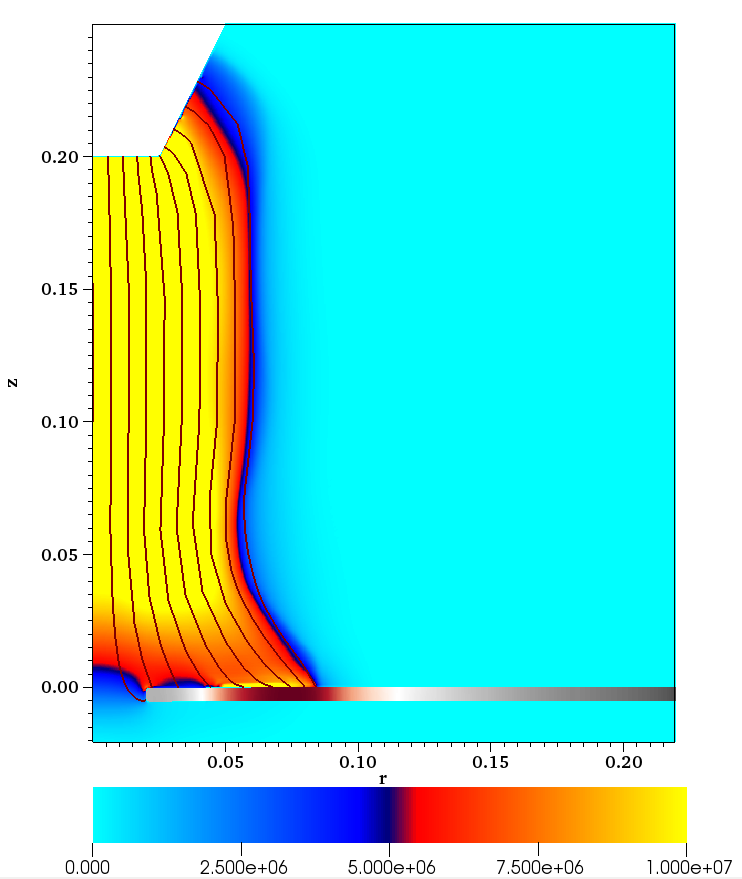}
\caption{low-conductivity substrate}
                \label{CFRPcurrentdensity}
        \end{subfigure}
\caption{The magnitude of the current density and current streamlines in the plasma arc with (left) an  aluminium substrate and (right) a low-conductivity substrate at $t=\SI{58.8}{\micro \second}$. The temperature in the substrate is plotted on the same scale for the two materials.}
\label{CFRPvsAL}
\end{figure}

Fig.\ \ref{CFRPvsAL} shows differences in the arc structure, with much
greater spreading at the arc root on the composite panel.
The lower electrical conductivity of the substrate is
  comparable to that of high-temperature plasma, thus the optimal
  current path is to travel further through the plasma, reducing the
  overall distance to the ground site.  Investigation of the
resulting reaction progress variable contours in the detonation showed
no difference between the two substrates hence we conclude that there
is no impact on detonation by the change of the substrate material.
This follows from the altered path of the current
  profile, the initial attachment provides the conditions to initiate
  the detonation.  However, the preference for the current streamlines
  to travel directly to the top of the substrate in both cases means
  additional energy input into the detonation is limited.

\subsection{Effect of a dielectric}

\begin{figure}[!tb]
\centering
\begin{subfigure}[b]{0.47\textwidth}
                \includegraphics[width=\textwidth]{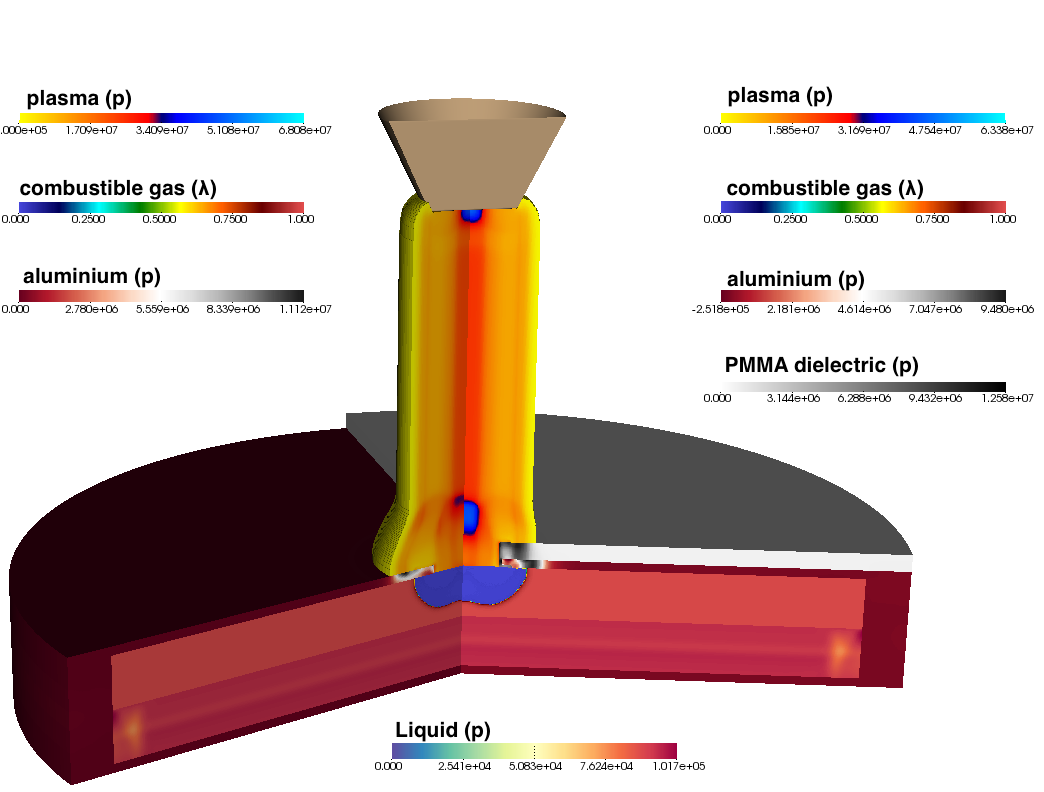}
\caption{The $\lambda$ field shows faster reaction (right) when the dielectric is present than when it is not (left), at $t=6.4{\micro \second}$. }
                \label{ALvsPMMA3D}
        \end{subfigure}\hspace{0.5cm}
        \begin{subfigure}[b]{0.42\textwidth}
                \includegraphics[width=\textwidth]{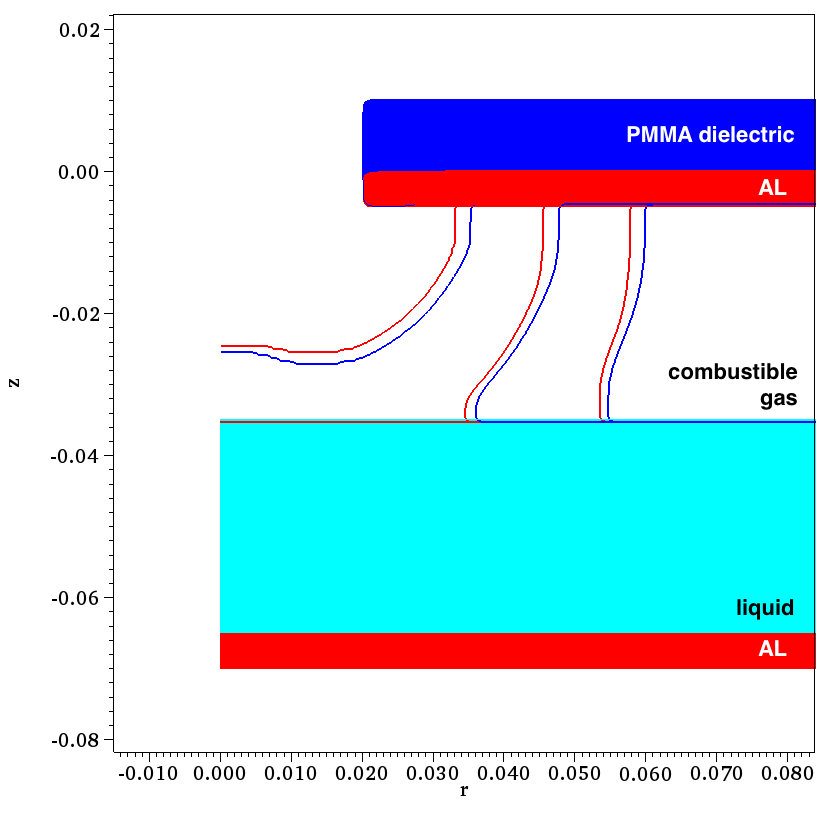}
\caption{The $\lambda=0.1$ contours at $t=6.4$, 11 and 16.8$\SI{}{\micro \second}.$}
                \label{ALvsPMMA2D}
        \end{subfigure}
\caption{The {\color{black} dielectric coating} placed over the {\color{black} aluminium} substrate  leads to a larger energy deposition in the combustible gas than the single {\color{black} aluminium} substrate layer. The shape of the plasma arc is also affected by the presence of the dielectric. In (b), the red line represents the reaction-front contour from the simulation of {\color{black} aluminium} substrate without dielectric and the blue line the contour from the simulation of {\color{black} aluminium} substrate with dielectric.}
\label{ALvsPMMA}
\end{figure}

We place a dielectric layer of PMMA, with negligible
  electrical conductivity
$2.6\times10^{-5}\SI{}{\siemens \per \meter}$ on top of the aluminium
substrate. The dielectric obeys a Mie-Gr\"uneisen equation of state as
given by equation (\ref{PMMAeos}) and parameters
(\ref{PMMAeos2}). Figures \ref{ALvsPMMA3D} and \ref{ALvsPMMA2D} show
that the dielectric layer, having a higher resistivity,
restricts the arc attachment area, thus we don't see
  the same expansion of the arc over the top surface of the
  substrate. Thus, the dielectric allows the deposition of more
energy directly to the combustible gas, leading to faster ignition of
the material compared to the  aluminium-only case.

\begin{figure}[!hbt]
\centering
\includegraphics[width=0.43\textwidth]{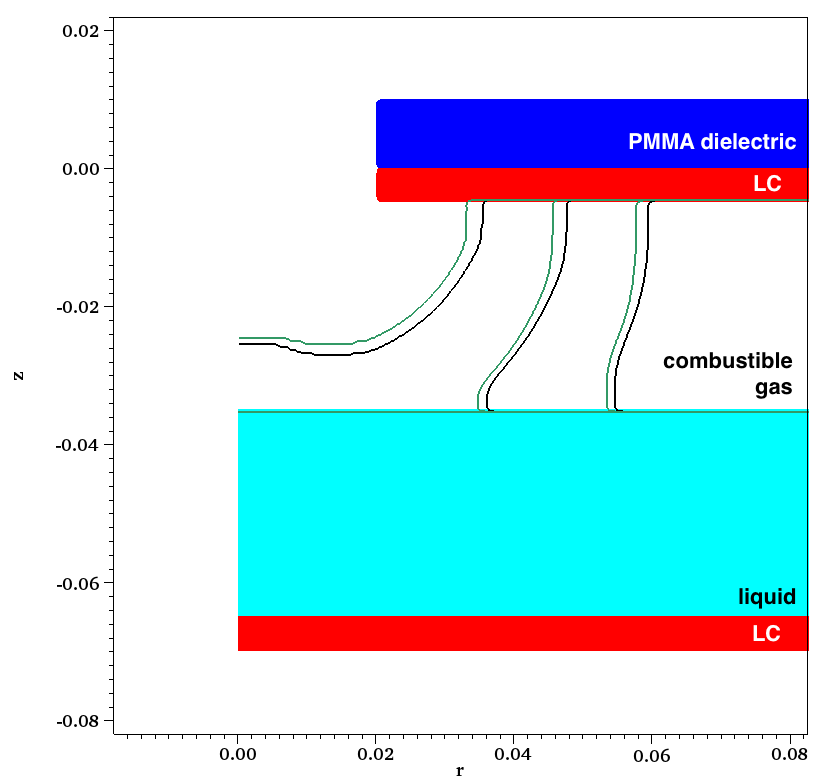}
\caption{The $\lambda=0.1$ contours at $t=6.4$, 11 and 16.8 $\SI{}{\micro \second}$, representing the reaction-front in the combustible gas. The green line represents the contour from the simulation of the LC substrate without dielectric and the black line the contour from the simulation of the LC substrate with dielectric.}
\label{CFRPvsPMMA}
\end{figure}

We repeat the same setup with a  low-conductivity
substrate, leading to the verification that the reaction is
accelerated by the effect of the dielectric, even when
aluminium is replaced by a 
  low-conductivity substrate. The respective $\lambda$ contours at
times $t=6.4$, 11 and  16.8 $\SI{}{\micro \second}$ are shown in
Figure \ref{CFRPvsPMMA}.

\subsection{Combustible gas sensitivity}

\begin{figure}[!hbt]
\centering
\includegraphics[width=0.43\textwidth]{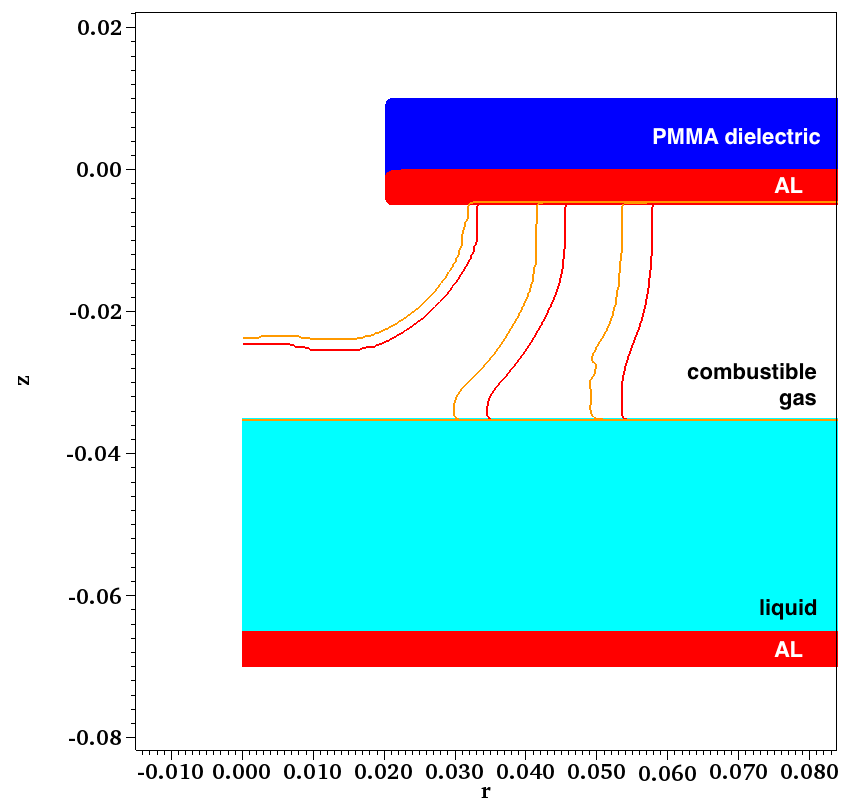}
\caption{The $\lambda=0.1$ contours at $t=6.4$, 11 and 16.8 $\SI{}{\micro \second}$ representing the reaction-front in the combustible gas. The red line represents the contour from the simulation of an  aluminium substrate (without dielectric) over a gas with activation temperature $T_A=8000\SI{}{\kelvin}$ and the orange line the contour from the simulation of an  aluminium substrate (without dielectric) over a gas with activation temperature $T_A=20000\SI{}{\kelvin}$. }
\label{8000vs20000}
\end{figure}

In the previous sections, the combustible gas had an activation
temperature of $8000\SI{}{\kelvin}$.  We alter the sensitivity
threshold of the combustible gas to $20000\SI{}{\kelvin}$ and
investigate the ignition process in this less sensitive combustible
material. The ignition is found to be slower in the less sensitive
material as expected, however, the  plasma is still
strong enough to ignite it directly with the initial strike. The
$\lambda=0.1$ contours are shown in Figure \ref{8000vs20000}. Both
simulations used an  aluminium substrate without a
dielectric.

\section{Conclusions}

In this work we present a methodology for the numerical simulation of the non-linear interaction of four states of matter; gas, liquid solid and plasma. In previous work we have considered the simultaneous solution of two or three states of matter and to the authors' knowledge this work is the first time the simultaneous solution of four states of matter is achieved. 
Each phase can be modelled by its own set of partial differential equations (PDEs) or multi-phase diffuse-interface models can be used to combine the PDEs of multiple phases. We present the coupling of our multi-phase, diffuse-interface formulation for plasma and gases with the compressible Euler formulation describing a liquid (or other fluid) material and the elastoplastic solid formulation in an Eulerian framework. The novelty of solving simultaneously the equations for the plasma arc and another material (e.g., solid) in the same  computational code should also be stressed. The basis of our algorithm is writing all the governing equations in the same, hyperbolic form, allowing all systems to be solved with (the same or distinct) finite-volume methods. To track the interfaces between materials described by different governing PDE systems, we utilise level set methods.  The communication between materials is achieved by solving mixed-material Riemann problems at the interfaces. To this end, we derive mixed Riemann solvers for each formulation pair considered in our application. These are based on the characteristic equations derived for each system pair, allowing for the computation of the star state in the `real' material, by taking into account the two different systems on either side of the interface and applying appropriate interface boundary conditions. In summary, our multi-physics methodology follows these steps: (a) Write all governing PDEs in the same mathematical form, (b) Solve each PDE system using finite volume, shock-capturing methods, (d) Use the level set method to track material interfaces (e) Use the Riemann ghost fluid method and the mixed-material Riemann solvers presented in this work for communication between materials.

We apply the new methodology to study scenarios of potential ignition
of combustible gases in a metal tank half-filled with liquid and
half-filled with combustible gas, which is  struck by lightning. 
We first considered a closed, cylindrical aluminium tank, half-filled with liquid and half-filled with gas, struck by lightning at the cylinder centreline. This allowed us to perform the simulations in cylindrical symmetry. It was proven in this case that, over the timescales of our simulations ($\sim 25\micro \second$), where heat conduction would not take place, the combustible material does not ignite. The pressure wave transmitted from the plasma through the substrate into the gas is weak, and the pressure does not built up enough over these timescales in the container to elevate the temperature more than a few degrees. Thus, the gas ignition threshold temperature is not reached in this scenario.
 We added a perforated dielectric layer over the aluminium substrate to direct the current over a smaller surface and potentially generate a funneling effect to examine whether that would lead to the ignition of the gas. Although the maximum pressures and temperatures reached in this scenario are higher than the uncoated aluminium scenario, they are still not enough to ignite the material.

 We also considered a perforated aluminium substrate. {\color{black}This would be akin to an improper lightning protection scheme that allowed the scenario where the lightning struck}
 and damaged the material and the same strike or another one hits the substrate at the
 same position. In this scenario, the mechanical impact of the
 lightning to the gas is direct and immediately ignites the
 gas. Replacing the aluminium with a lower-conductivity material did
 not affect the ignition; the gas ignited immediately again and the
 combustion process continued with the same rate.

We added a perforated dielectric layer over the perforated metal substrate and concluded that the presence of the dielectric accelerates the combustion of the gas, both for an aluminium and a lower-conductivity substrate. 

Finally, replacing the combustible gas with a much less sensitive gaseous material demonstrated the deceleration of the combustion process, although this was not inhibited; the lightning strike still  ignited the  material directly. 

The main factor that inhibited the combustion of the gaseous material was the sealing of the top surface of the metal container, suggesting that extra care should be taken in cases where materials have been pre-damaged. Thus, our methodology can be used as part of the manufacturing process for optimising  compartments in terms of shapes and materials.

\section*{Acknowledgements}

This work was supported by Jaguar Land Rover and the UK-EPSRC grant EP/K014188/1
as part of the jointly funded Programme for Simulation Innovation and Boeing
Research \& Technology (BR\&T) grant SSOW-BRT-L0516-0569. The authors gratefully thank Alan Minchinton from Orica - Research \& Innovation for useful discussions.

\bibliographystyle{elsarticle-num}
\bibliography{refs,lightning}

\end{document}